\documentclass[aps,prb,twocolumn,superscriptaddress]{revtex4-1}
\usepackage[utf8]{inputenc}
\usepackage{amsfonts,amsmath,amssymb,amsthm} 
\usepackage{wasysym} 
\usepackage{bbold}
\usepackage[usenames,dvipsnames]{xcolor}
\usepackage[pdftex]{graphicx}
\usepackage{bm}
\usepackage[pdftex]{hyperref}
{\end{pmatrix}\end{medsize}} 
\usepackage{nccmath} 
\usepackage{natbib} 
\usepackage{physics} 
\usepackage{ulem}

\hypersetup{colorlinks=true,linkcolor=red, citecolor=blue, urlcolor=blue,colorlinks=true,bookmarks=true,breaklinks=true}

\usepackage{array}
\newcolumntype{L}[1]{>{\raggedright\let\newline\\\arraybackslash\hspace{0pt}}m{#1}}
\newcolumntype{C}[1]{>{\centering\let\newline\\\arraybackslash\hspace{0pt}}m{#1}}
\newcolumntype{R}[1]{>{\raggedleft\let\newline\\\arraybackslash\hspace{0pt}}m{#1}}

\begin{document}
\title{Non-equilibrium evolution of the optical conductivity of the weakly interacting Hubbard model: Drude response and $\pi$-ton type vertex corrections}
\author{Olivier \surname{Simard}}
\affiliation{Department of Physics, University of Fribourg, 1700 Fribourg, Switzerland}
\author{Martin \surname{Eckstein}}
\affiliation{Department of Physics, University of Erlangen-Nürnberg, 91058 Erlangen, Germany}
\author{Philipp \surname{Werner}}
\affiliation{Department of Physics, University of Fribourg, 1700 Fribourg, Switzerland}
\date{\today}

\begin{abstract}
The optical conductivity contains information about energy absorption and the underlying physical processes. In finite-dimensional systems, vertex corrections to the bare bubble need to be considered, which is a computationally challenging task. Recent numerical studies showed that in the weak coupling limit, near an ordering instability with wave vector $\pi$, the vertical ladder describing particle-hole pairs interacting via the exchange of this wave vector becomes the dominant vertex correction. The corresponding Maki-Thompson-like diagram has been dubbed $\pi$-ton. Here, we add the $\pi$-ton ladder vertex correction to dynamical mean field theory estimates of the optical conductivity.
By performing calculations on the Kadanoff-Baym contour, we reveal the characteristic spectral signatures of the $\pi$-tons and their evolution under non-equilibrium conditions. We consider interaction quenches of the weakly-correlated Hubbard model near the antiferromagnetic phase boundary, and analyze the evolution of the Drude and $\pi$-ton features. While the bubble contribution to the optical conductivity is found to thermalize rapidly, after some oscillations with frequencies related to the local spectral function, the $\pi$-ton contribution exhibits a slower evolution. We link this observation to the prethermalization phenomenon which has been previously studied in weakly interacting, quenched Hubbard models.  
\end{abstract}
\maketitle


\section{Introduction}
\label{sec:Introduction}

Computing the electronic structure and transport properties of non-equilibrium correlated electron systems is a challenging task. In low-dimensional lattice systems, one-particle quantities such as the spectral function or self-energy are influenced by two-particle correlation functions. This is especially the case close to ordering instabilities where irreducible vertices depend strongly on momentum and non-local correlations need to be taken into account.\cite{Rohringer2018,Bergeron_2011_optical_cond}
For example, low-energy spin and charge correlations  can leave clear signatures in the electronic quasi-particle spectra, \citep{kusunose_influence_2006,RevModPhys.77.1027,Rohringer2018} and the optical conductivity and related susceptibilities are strongly modified by vertex corrections to the leading particle-hole contribution. 

Recent investigations\citep{kauch_pitons_2019,worm2020broadening,PhysRevB.103.104415_Simard_pi_ton} in Hubbard-type models have shown that in the vicinity of a charge density wave (CDW) or antiferromagnetic (AFM) instability, the dominant vertex corrections in transport quantities, such as the longitudinal optical conductivity, stem from a vertical ladder that exchanges momentum $\mathbf{k}-\mathbf{k}^{\prime}\simeq (\pi,\pi,\cdots) \equiv \mathbf{k}_{\pi}$.\citep{Hubbard_1963} 
This vertical ladder vertex correction, dubbed $\pi$-ton,\cite{kauch_pitons_2019} describes physical processes in which a particle-hole pair creates another particle-hole pair at a wave vector $\mathbf{k}_{\pi}$, and these interact with each other until recombination occurs.
Since $\pi$-tons are spectral features which grow significantly as the system approaches the ordering instability,\cite{kauch_pitons_2019,worm2020broadening,PhysRevB.103.104415_Simard_pi_ton} they allow to track the relevant correlations in the precursor state to the ordered phase, and it is thus interesting to also study these features in out-of-equilibrium situations. Nonthermal transient enhancements of the spin susceptibility\cite{PhysRevB.92.024305_Bauer_short_range_order} or pairing susceptibility\cite{Stahl_2021} have been previously discussed in theoretical works which considered interaction quenches starting from the disordered phase. In the weak-coupling regime, such quench dynamics can be influenced by trapping phenomena\cite{Eckstein_2009,Tsuji_2013} related to prethermalization\citep{Moeckel_2008} or nonthermal fixed points.\citep{Berges_2004,tsuji_nonequilibrium_2013} Experimentally, interaction quenches can be realized in cold-atom systems, where the antiferromagnetic phase of the Fermi-Hubbard model has recently been accessed.\cite{Mazurenko_2017} In these systems, spin correlations can be detected directly by means of quantum gas microscopy.\cite{Edge_2015,Mazurenko_2017}  

In Ref.~\onlinecite{PhysRevB.103.104415_Simard_pi_ton}, it was demonstrated that the $\pi$-ton-type vertex corrections to the spin susceptibility and optical conductivity of the half-filled Hubbard model can be qualitatively captured by a post-processing analysis of dynamical mean field theory (DMFT)\cite{Georges_1996} data. This Random Phase Approximation (RPA) $\pi$-ton approach is expected to work at weak coupling and near the ordering instability, where the single-ladder  vertex correction is dominant.\cite{PhysRevB.103.104415_Simard_pi_ton}  The latter study however only considered systems in equilibrium, using a Matsubara formalism. To more clearly reveal the $\pi$-ton signatures in the optical conductivity and in related susceptibilities, and to study the evolution of these spectral features under non-equilibrium conditions, we evaluate here the correlation functions and spectra using real-time simulations based on the NESSi library.\cite{Nessi} 

We will use the real-time formalism to analyze the $\pi$-ton and Drude response during and after interaction ramps and interaction quenches in the vicinity of the antiferromagnetically ordered phase of the half-filled, weakly interacting Hubbard model. 
This investigation reveals significantly different timescales for the relaxation, prethermalization and thermalization of the Drude peak and the spectral feature associated with the $\pi$-ton. 

The paper is structured as follows. The Hubbard model and method employed to solve the non-equilibrium DMFT equations are presented in Sec.~\ref{sec:Models_and_methods}. In Sec.~\ref{subsec:susceptibilities} we present the formulae for the RPA-ladder-type $\pi$-ton vertex corrections. The non-equilibrium diagrammatic results for the half-filled Hubbard model are presented in Sec.~\ref{sec:results}. The discussion and conclusions can be found in Secs.~\ref{sec:discussion} and \ref{sec:conclusion}, respectively.


\section{Model and method}
\label{sec:Models_and_methods}

\subsection{Hubbard model}
\label{subsec:Hubbard_model}

We consider a single-band Hubbard model on a $D$-dimensional hypercubic lattice  with a time-dependent interaction parameter
\begin{align}
\label{eq:Hubbard_model_intro}
\hat{\mathcal{H}}(t)=&-t_\text{hop}\sum_{\langle i,j\rangle,\sigma}\left(\hat{c}_{i,\sigma}^{\dagger}\hat{c}_{j,\sigma}+\text{H.c.}\right)+U(t)\sum_i\hat{n}_{i,\uparrow}\hat{n}_{i,\downarrow}\nonumber\\
&-\mu \sum_i (\hat{n}_{i,\uparrow} + \hat{n}_{i,\downarrow}).
\end{align} 
Here, $t_\text{hop}$ is the nearest-neighbor hopping amplitude, $\langle i,j\rangle$ denotes nearest-neighbor pairs, $\sigma \in \{\uparrow,\downarrow\}$ denotes the spin, and $\hat{c}^{(\dagger)}_{i,\sigma}$ the annihilation (creation) operators for site $i$. Furthermore, $\hat{n}_{i\sigma}=\hat{c}^{\dagger}_{i,\sigma}\hat{c}_{i,\sigma}$ is the number operator, $U(t)$ is the time-dependent local Hubbard repulsion and $\mu$ the chemical potential. We use $t_\text{hop}$ as the unit of energy and $\hbar/t_\text{hop}$ as the unit of time. We set $\hbar$, $k_B$, the electric charge $e$ and the lattice spacings $a$ equal to unity. All the calculations will be for half-filled systems with $\mu=U/2$.

\subsection{Non-equilibrium DMFT}
\label{subsec:DMFT}

\subsubsection{General formalism}

\begin{figure}[t]
  \centering
    \includegraphics[width=1.0\linewidth]{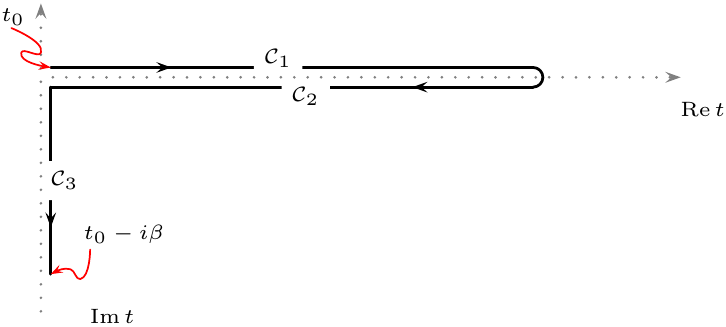}
      \caption{Illustration of the Kadanoff-Baym coutour. This contour starts off at time $t_0$ and goes to some time $t_{\text{max}}$, returns to $t_0$, and then extends along the imaginary-time axis to $t_0-i\beta$, where $\beta$ is the inverse temperature of the initial equilibrium state. The arrows indicate the contour ordering.
      }
  \label{fig:kadanoff_baym_contour}
\end{figure}

Non-equilibrium DMFT is an implementation of the DMFT equations on the Kadanoff-Baym contour $\mathcal{C}$ (see Fig.~\ref{fig:kadanoff_baym_contour}).\cite{RevModPhys.86.779_non_eq_review,PhysRevLett.97.266408_freericks_non_eq} DMFT is based on the assumption of a local self-energy, which becomes exact in infinite-dimensional lattices.\cite{PhysRevLett.62.324,Georges_1992,Mueller_1989} Even in low dimensions (including the $D=1$ case considered below) DMFT yields a solution that is characteristic of high-dimensional lattices. In particular, it produces an equilibrium phase diagram with a nonzero N\'eel temperature for $U>0$ and half-filling. We will study the nonequilibrium properties of $\pi$-ton type vertex corrections for interaction ramps and quenches in the vicinity of this ordering transition, by adapting the formalism introduced in Ref.~\onlinecite{PhysRevB.103.104415_Simard_pi_ton}.

In DMFT, the lattice model is self-consistently mapped onto a single-site impurity model, where upon convergence the time-dependent hybridization function represents the effects of the lattice environment.\cite{Georges_1996} The action of the non-equilibrium impurity problem can be written as
\begin{align}
\label{eq:DMFT:DMFT_action}
\mathcal{S}[\Delta] =& -\int_{\mathcal{C}}\mathrm{d}z \ \hat{\mathcal{H}}_{\text{loc}}(z) \nonumber\\
& - \int_{\mathcal{C}}\mathrm{d}z \int_{\mathcal{C}} \mathrm{d}z^{\prime} \sum_\sigma \hat{c}^{\dagger}_\sigma(z)\Delta^\sigma(z,z^{\prime})\hat{c}_\sigma(z^{\prime}),
\end{align} 
where $ \hat{\mathcal{H}}_{\text{loc}}$ is the same local term as in the lattice model, $\hat{c}^{(\dagger)}_\sigma$ annihilates (creates) an electron with spin $\sigma$ on the impurity and $z\in \mathcal{C}\equiv \mathcal{C}_1\oplus\mathcal{C}_2\oplus\mathcal{C}_3$. The hybridization function is denoted by $\Delta^\sigma(z,z^{\prime})$, and the integral is over the Kadanoff-Baym contour $\mathcal{C}$, which is represented in Fig.~\ref{fig:kadanoff_baym_contour} with the forward real-time branch $\mathcal{C}_1$, the backward real-time branch $\mathcal{C}_2$, and the vertical imaginary-time branch $\mathcal{C}_3$.

With the non-equilibrium action \eqref{eq:DMFT:DMFT_action}, one can define the non-equilibrium impurity Green's function

\begin{align}
\label{eq:DMFT:Greens_function}
\mathcal{G}^{\sigma}(z,z^{\prime}) = -i\text{Tr}\left[\mathcal{T}_{\mathcal{C}}e^{i\mathcal{S}[\Delta]}\hat{c}_{\sigma}(z)\hat{c}^{\dagger}_\sigma(z^{\prime})\right]/\mathcal{Z},
\end{align}
where $\mathcal{T}_{\mathcal{C}}$ is the time-ordering operator defined on the Kadanoff-Baym contour (following the arrows in Fig.~\ref{fig:kadanoff_baym_contour}) and $\mathcal{Z}$ is the partition function $\mathcal{Z} = \text{Tr}\left[\mathcal{T}_{\mathcal{C}}e^{i\mathcal{S}[\Delta]}\right]$. The Green's function \eqref{eq:DMFT:Greens_function} and all objects defined on the Kadanoff-Baym contour can be represented in a matrix-form, consisting of a $3\times 3$ matrix holding all the combinations of the three components composing $\mathcal{C}$:

\begin{align}
\label{eq:non_eq_kadanoff_baym}
\underline{\mathcal{G}}^{\sigma}(z,z^{\prime})=\begin{pmatrix}
\mathcal{G}_{11}^{\sigma}(z,z^{\prime})&\mathcal{G}_{12}^{\sigma}(z,z^{\prime})&\mathcal{G}_{13}^{\sigma}(z,z^{\prime})\\
\mathcal{G}_{21}^{\sigma}(z,z^{\prime})&\mathcal{G}_{22}^{\sigma}(z,z^{\prime})&\mathcal{G}_{23}^{\sigma}(z,z^{\prime})\\
\mathcal{G}_{31}^{\sigma}(z,z^{\prime})&\mathcal{G}_{32}^{\sigma}(z,z^{\prime})&\mathcal{G}_{33}^{\sigma}(z,z^{\prime})
\end{pmatrix}.
\end{align}
Here, $\mathcal{G}^{\sigma}_{\alpha\beta}$ means that the annihilation operator is on branch $\mathcal{C}_\alpha$ and the creation operator on $\mathcal{C}_\beta$.
To represent the matrix objects, we make use of the underline bar as on the left hand side of Eq.~\eqref{eq:non_eq_kadanoff_baym}. However, not all elements of Eq.~\eqref{eq:non_eq_kadanoff_baym} are independent and four elements are sufficient to reconstruct the matrix. We choose 
the retarded, lesser, left-mixing and Matsubara components:\cite{RevModPhys.86.779_non_eq_review,stefanucci_van_leeuwen_2013}
\begin{align}
\label{eq:non_eq_components_cases}
\begin{cases}
\begin{aligned}
 \mathcal{G}^{R,\sigma}(t,t^{\prime}) & = \frac{1}{2}\left(\mathcal{G}^{\sigma}_{11}+\mathcal{G}^{\sigma}_{21}-\mathcal{G}^{\sigma}_{12}-\mathcal{G}^{\sigma}_{22}\right)\\
  &= -i\Theta(t-t^{\prime})\langle\{\hat{c}_{\sigma}(t),\hat{c}^{\dagger}_{\sigma}(t^{\prime})\}\rangle,\\
 \mathcal{G}^{<,\sigma}(t,t^{\prime}) & = \mathcal{G}^{\sigma}_{12} = i\langle\hat{c}_{\sigma}^{\dagger}(t^{\prime})\hat{c}_{\sigma}(t)\rangle,\\
 \mathcal{G}^{\invneg,\sigma}(\tau,t^{\prime}) &= \frac{1}{2}\left(\mathcal{G}^{\sigma}_{31}+\mathcal{G}^{\sigma}_{32}\right) = -i\langle\hat{c}_{\sigma}(\tau)\hat{c}_{\sigma}^{\dagger}(t^{\prime})\rangle,\\
 \mathcal{G}^{\text{M},\sigma}(\tau,\tau^{\prime}) &= -i\mathcal{G}^{\sigma}_{33} = -\langle\mathcal{T}_{\tau}\hat{c}_{\sigma}(\tau)\hat{c}_{\sigma}^{\dagger}(\tau^{\prime})\rangle.
 \end{aligned}
\end{cases}
\end{align}
Real times will be denoted by latin letters $t\in \mathcal{C}_1\cup\mathcal{C}_2$, and imaginary time by Greek letters $\tau\in \mathcal{C}_3$ (see Fig.~\ref{fig:kadanoff_baym_contour}).

Similarly as in Ref.~\onlinecite{PhysRevB.103.104415_Simard_pi_ton}, the impurity Green's function will be computed using the iterated perturbation theory~\citep{kajueter_new_1996,tsuji_nonequilibrium_2013} (IPT) method, adapted to the non-equilibrium formalism (see Sec.~\ref{subsubsec:IPT}). 
The longitudinal optical conductivity and its vertex corrections are computed using the RPA post-processing method described in Ref.~\onlinecite{PhysRevB.103.104415_Simard_pi_ton}. The corresponding non-equilibrium equations are presented in Sec.~\ref{subsec:susceptibilities}.

\subsubsection{Paramagnetic self-consistency}
\label{subsubsec:PM}

The self-consistency condition demands that the impurity Green's function 
$\underline{\mathcal{G}}^{\sigma}(z,z^{\prime})$ is identical to the local lattice Green's function $\underline{\mathcal{G}}^{\sigma}_\text{loc}(z,z^{\prime})$. 
This self-consistency condition implicitly fixes the hybridization function $\underline{\Delta}^\sigma(z,z')$ which is needed in the impurity action Eq.~\eqref{eq:DMFT:DMFT_action}. This hybridization function plays the role of a dynamical mean field. Alternatively, one can define a so-called Weiss Green's function $\underline{\mathcal{G}}^{\sigma}_{0}$, which is related to the hybridization function by\cite{Georges_1996} 
\begin{align}
\label{eq:PM:Weiss_Green_hyb}
\left[i\partial_z+\mu\right]\underline{\mathcal{G}}^{\sigma}_0(z,z^{\prime})-\int_{\mathcal{C}}\mathrm{d}\bar{z} \ \underline{\Delta}^{\sigma}(z,\bar{z})\underline{\mathcal{G}}_0^{\sigma}(\bar{z},z^{\prime})=\delta^{\mathcal{C}}(z,z^{\prime}),
\end{align} 
where $\delta^{\mathcal{C}}(z,z')$ represents the delta function on the Kadanoff-Baym contour. The impurity Dyson equation links the Weiss Green's function $\underline{\mathcal{G}}_0^{\sigma}$, the impurity Green's function $\underline{\mathcal{G}}^{\sigma}$, and the impurity self-energy $\underline{\Sigma}^{\sigma}$: 

\begin{align}
\label{eq:PM:Dyson_eq}
&\underline{\mathcal{G}}^{\sigma}(z,z^{\prime}) =\notag\\
&\phantom{=}\underline{\mathcal{G}}_{0}^{\sigma}(z,z^{\prime}) + \int_{\mathcal{C}}\mathrm{d}\bar{z}\int_{\mathcal{C}}\mathrm{d}\bar{z}^{\prime}\underline{\mathcal{G}}_{0}^{\sigma}(z,\bar{z})\underline{\Sigma}^{\sigma}(\bar{z},\bar{z}^{\prime})\underline{\mathcal{G}}^{\sigma}(\bar{z}^{\prime},z^{\prime}).
\end{align}
In non-equilibrium DMFT, the lattice self-energy is set equal to the local impurity self-energy, $\underline{\Sigma}^\sigma_{ij}(z,z^{\prime})=\underline{\Sigma}^\sigma(z,z^{\prime})\delta_{ij}$, which is an approximation in finite-dimensional systems.\cite{Mueller_1989} Both $\underline{\Sigma}^{\sigma}[\underline{\mathcal{G}}_0^{\sigma}]$ and $\underline{\Sigma}^{\sigma}[\underline{\mathcal{G}}^{\sigma}]$ will be computed using the impurity solvers described in Sec.~\ref{subsubsec:IPT}.

With the DMFT approximation on the self-energy, the Dyson equation for the lattice Green's function $\underline{\mathcal{G}}_{\mathbf{k}}^{\sigma}$ can be written as
\begin{align}
\label{eq:PM:projected_green_function_impurity}
&\left[i\partial_z+\mu-\epsilon(\mathbf{k})\right]\underline{\mathcal{G}}_{\mathbf{k}}^{\sigma}(z,z^{\prime})-\int_{\mathcal{C}}\mathrm{d}\bar{z} \ \underline{\Sigma}^{\sigma}(z,\bar{z})\underline{\mathcal{G}}_{\mathbf{k}}^{\sigma}(\bar{z},z^{\prime})\notag\\
&=\delta^{\mathcal{C}}(z,z^{\prime}),
\end{align}
where $\epsilon(\mathbf{k})$ is the bare electronic dispersion.
Reshuffling Eq.~\eqref{eq:PM:Dyson_eq} and substituting the impurity $\underline{\mathcal{G}}^{\sigma}$ by the $\mathbf{k}$-averaged lattice Green's function $\underline{\mathcal{G}}_\text{loc}^{\sigma}$, one can obtain the following Volterra integral equation
\begin{align}
\label{eq:PM:impurity_self_energy}
\int_{\mathcal{C}}\mathrm{d}\bar{z} \ \underline{\mathcal{G}}_0^{\sigma}(z,\bar{z})\left[\delta^{\mathcal{C}}(\bar{z},z^{\prime}) + \underline{F}^{\sigma}(\bar{z},z^{\prime})\right] = \underline{\mathcal{G}}^{\sigma}_\text{loc}(z,z^{\prime}),
\end{align} where $\underline{F}^{\sigma}(\bar{z},z^{\prime})\equiv \int_{\mathcal{C}}\mathrm{d}\bar{z}^{\prime} \ \underline{\Sigma}^{\sigma}(\bar{z},\bar{z}^{\prime})\underline{\mathcal{G}}^{\sigma}_\text{loc}(\bar{z}^{\prime},z^{\prime})$. Equations~\eqref{eq:PM:projected_green_function_impurity} and \eqref{eq:PM:impurity_self_energy} form, along with the IPT expression for the impurity self-energy, a closed set of equations which determines $\underline{\mathcal{G}}_0^{\sigma}$.\cite{Eckstein2010ipt,tsuji_nonequilibrium_2013} We can directly insert the IPT self-energy 
into Eq.~\eqref{eq:PM:projected_green_function_impurity} and the impurity Dyson equation and iterate the solution until convergence. To solve the Dyson equations \eqref{eq:PM:Dyson_eq}, \eqref{eq:PM:projected_green_function_impurity} and Volterra integral equation \eqref{eq:PM:impurity_self_energy}, the NESSi package is used.\cite{Nessi} For the paramagnetic (PM) solution, we impose $\underline{\Sigma}^\uparrow=\underline{\Sigma}^\downarrow$ and similarly for $\underline{\mathcal{G}}^{\sigma}$ and $\underline{\Delta}^{\sigma}$.

\subsubsection{IPT solver}
\label{subsubsec:IPT} 
Since we work in the weak coupling regime ($U \lesssim \text{bandwidth}/2$), we use IPT as impurity solver. IPT is a second-order perturbation theory for the Anderson impurity model.\cite{kajueter_new_1996,arsenault_benchmark_2012,tsuji_nonequilibrium_2013} In the ``bare IPT" formalism, the self-energy is approximated as 
\begin{align}
\label{eq:PM:impurity_self_energy_IPT}
\underline{\Sigma}^{\sigma}[\underline{\mathcal{G}}_0^{\sigma}](z,z^{\prime}) = U(z)U(z^{\prime})\underline{\mathcal{G}}^{\sigma}_{0}(z,z^{\prime})\underline{\mathcal{G}}^{-\sigma}_{0}(z,z^{\prime})\underline{\mathcal{G}}^{-\sigma}_{0}(z^{\prime},z),
\end{align} 
and hence is a functional of the Weiss Green's function defined in Eq.~\eqref{eq:PM:Weiss_Green_hyb}. The interaction $U(z)$ is a function on the contour $\mathcal{C}$ that relates to $U(t)$ in Eq.~(\ref{eq:Hubbard_model_intro}) in the following way: on $\mathcal{C}_3$ its value is $U(z\in\mathcal{C}_3)=U(t=0^-)$, namely the interaction of the initial equilibrium state, whereas on the real-time branches $\mathcal{C}_1$ and $\mathcal{C}_2$ it corresponds to $U(z\in\mathcal{C}_1\oplus\mathcal{C}_2)=U(t)$, where $t\ge 0$ is the time associated with $z$. Note that at half-filling, by choosing $\mu=U/2$, the Hartree term vanishes in the paramagnetic state. 

Alternatively, one can define a ``bold IPT," where $\underline{\mathcal{G}}_0^{\sigma}$ is replaced by the dressed impurity Green's function $\underline{\mathcal{G}}^{\sigma}$ in Eq.~\eqref{eq:PM:impurity_self_energy_IPT}. This replacement does not severely affect the results for $U\lesssim \text{bandwith}/2$, but it yields a conserving approximation, which means that the total energy after a perturbation is conserved under the time evolution.\citep{Eckstein2010ipt}

\subsection{Susceptibilities}
\label{subsec:susceptibilities}

\begin{figure}[t]
  \centering
    \includegraphics[width=1.0\linewidth]{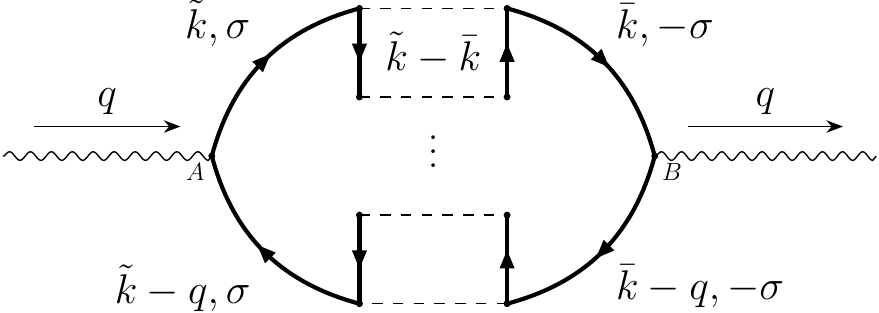}
      \caption{Illustration of the single-ladder vertex correction to the susceptibilities. All diagrams sharing this topology are summed up in Eq.~\eqref{eq:sus:non_eq_sl_KB_vertx_corr}. To obtain the conductivity $\chi_{j_ij_i}$, the vertices $A$ and $B$ are both set equal to the velocity $v_i$.
      }
  \label{fig:single_ladder_diagrams_corr_sus}
\end{figure}

On the Kadanoff-Baym contour, the general expression for the single-ladder vertex corrections to the longitudinal optical conductivity, illustrated in Fig.~\ref{fig:single_ladder_diagrams_corr_sus}, can be computed from
\begin{align}
\label{eq:sus:non_eq_sl_KB_vertx_corr}
&\underline{\chi}_{\text{sl},j_ij_i}^{\sigma,-\sigma}(\mathbf{q}; z,z^{\prime}) = \notag\\
&\phantom{0}-\int_{-\pi}^{\pi}\frac{\mathrm{d}^{D}\tilde{k}}{(2\pi)^{D}}\int_{-\pi}^{\pi}\frac{\mathrm{d}^{D}\bar{k}}{(2\pi)^{D}}\int_{\mathcal{C}}\mathrm{d}\bar{z}\int_{\mathcal{C}}\mathrm{d}\bar{z}^{\prime}v_i(\tilde{\mathbf{k}})v_i(\bar{\mathbf{k}})\times\notag\\
&\phantom{0}\underline{\mathcal{G}}^{\sigma}_{\tilde{\mathbf{k}}}(z,\bar{z})\underline{\mathcal{G}}^{\sigma}_{\tilde{\mathbf{k}}-\mathbf{q}}(\bar{z}^{\prime},z)\underline{\square}^{\sigma,-\sigma}_{\tilde{\mathbf{k}}-\bar{\mathbf{k}}}(\bar{z},\bar{z}^{\prime})\underline{\mathcal{G}}^{-\sigma}_{\bar{\mathbf{k}}}(\bar{z},z^{\prime})\underline{\mathcal{G}}^{-\sigma}_{\bar{\mathbf{k}}-\mathbf{q}}(z^{\prime},\bar{z}^{\prime}),
\end{align}
by multiplying at the times $z$ and $z'$ with the velocities $v_i(\mathbf{k})=\partial_{k_i}\epsilon(\mathbf{k})$.\cite{PhysRevB.103.104415_Simard_pi_ton} Here the subscript $i$ specifies a Cartesian axis in 1D. The box ``$\underline{\square}$'' represents the vertical ladder vertex corrections and will be detailed below. Using Eqs.~\eqref{eq:non_eq_kadanoff_baym} and \eqref{eq:non_eq_components_cases}, as well as the Langreth rules,\cite{stefanucci_van_leeuwen_2013} the nine components of the $\underline{\chi}_{\text{sl}}$ matrix can be written down and numerically evaluated. In the following, we only consider the case where $\mathbf{q}=\mathbf{0}$ (optical excitations with low-energy photons), and we only need the lesser and greater components, denoted as $\chi^{<}_{\text{sl}}$ and $\chi^{>}_{\text{sl}}$, respectively. Therefore, we calculate the solution for variables $z$ and $z^{\prime}$ on the real-time branches $\mathcal{C}_1$ and $\mathcal{C}_2$ (for details, see Appendix~\ref{sec:appendice:sl_vertex_corr_on_KB_contour}).

The vertical ladder denoted by ``$\underline{\square}$'' is the solution of a singular Volterra integral equation that needs to be obtained before attaching the four outer Green's functions as in Eq.~\eqref{eq:sus:non_eq_sl_KB_vertx_corr}. It can be decomposed into the terms $\square^{\delta}(z)\delta^{\mathcal{C}}(z,z^{\prime})+\square^<(z,z^{\prime})\theta^{\mathcal{C}}(z^{\prime},z)+\square^>(z,z^{\prime})\theta^{\mathcal{C}}(z,z^{\prime})$, where $\square^{\delta}(z) = U(z)$ and $\theta^\mathcal{C}(z,z')$ is the Heaviside function on $\mathcal{C}$.

The integral equation for the ladder reads
\begin{align}
\label{eq:sus:vertical_ladder_sin_vie2}
&\underline{\square}_{\tilde{\mathbf{k}}-\bar{\mathbf{k}}}^{\sigma,-\sigma}(z,z^{\prime}) = U(z)\delta^{\mathcal{C}}(z,z^{\prime}) -\notag\\
&\phantom{0}U(z)\int_{-\pi}^{\pi}\frac{\mathrm{d}^Dk}{(2\pi)^{D}}\int_{\mathcal{C}}\mathrm{d}\bar{z} \ \underline{\mathcal{G}}^{\sigma}_{\mathbf{k}}(z,\bar{z})\underline{\mathcal{G}}^{-\sigma}_{\mathbf{k}-\tilde{\mathbf{k}}+\bar{\mathbf{k}}}(\bar{z},z)\underline{\square}_{\tilde{\mathbf{k}}-\bar{\mathbf{k}}}^{\sigma,-\sigma}(\bar{z},z^{\prime}),
\end{align} 
or, after some reshuffling,
\begin{align}
\label{eq:sus:vertical_ladder_sin_vie2_explicited}
&\int_{\mathcal{C}}\mathrm{d}\bar{z}\biggl[ \delta^{\mathcal{C}}(z,\bar{z}) + U(z)\int_{-\pi}^{\pi}\frac{\mathrm{d}^Dk}{(2\pi)^{D}}\underline{\mathcal{G}}^{\sigma}_{\mathbf{k}}(z,\bar{z})\underline{\mathcal{G}}^{-\sigma}_{\mathbf{k}-\tilde{\mathbf{k}}+\bar{\mathbf{k}}}(\bar{z},z)\biggr]\times\notag\\
&\phantom{0}\underline{\square}_{\tilde{\mathbf{k}}-\bar{\mathbf{k}}}^{\sigma,-\sigma}(\bar{z},z^{\prime}) = U(z)\delta^{\mathcal{C}}(z,z^{\prime}).
\end{align}
Since we are limited in memory, we approximate the ${\bf k}$ integrals in Eqs.~\eqref{eq:sus:non_eq_sl_KB_vertx_corr}-\eqref{eq:sus:vertical_ladder_sin_vie2_explicited} by a Riemann sum over 34 ${\bf k}$ points. With this number of ${\bf k}$ points, measured quantities are converged in the parameter regime considered.

The total longitudinal optical conductivity $\underline{\chi}_{j_ij_i\mathbf{q}}(z,z^{\prime})$ is obtained by adding the single-ladder correction \eqref{eq:sus:non_eq_sl_KB_vertx_corr} (with velocity factors) to the bare bubble
\begin{align}
\label{eq:sus:non_interacting_bubble}
&\underline{\chi}^0_{j_ij_i\mathbf{q}}(z,z^{\prime}) = \notag\\
&\phantom{=} -2\int_{-\pi}^{\pi}\frac{\mathrm{d}^Dk}{(2\pi)^D} \ v_i(\mathbf{k})v_i(\mathbf{k}+\mathbf{q})\underline{\mathcal{G}}_{\mathbf{k}}(z,z^{\prime})\underline{\mathcal{G}}_{\mathbf{k}+\mathbf{q}}(z^{\prime},z),
\end{align} namely $\underline{\chi}_{j_ij_i\mathbf{q}} = \underline{\chi}^0_{j_ij_i\mathbf{q}} +\underline{\chi}_{\text{sl},j_ij_i\mathbf{q}}$. In Eq.~\eqref{eq:sus:non_interacting_bubble}, the factor of 2 comes from the trace over the spin degrees of freedom. In the rest of the paper, we do not explicitly write the spin index.

  
\begin{figure}[b]
  \centering
    \includegraphics[width=1.0\linewidth]{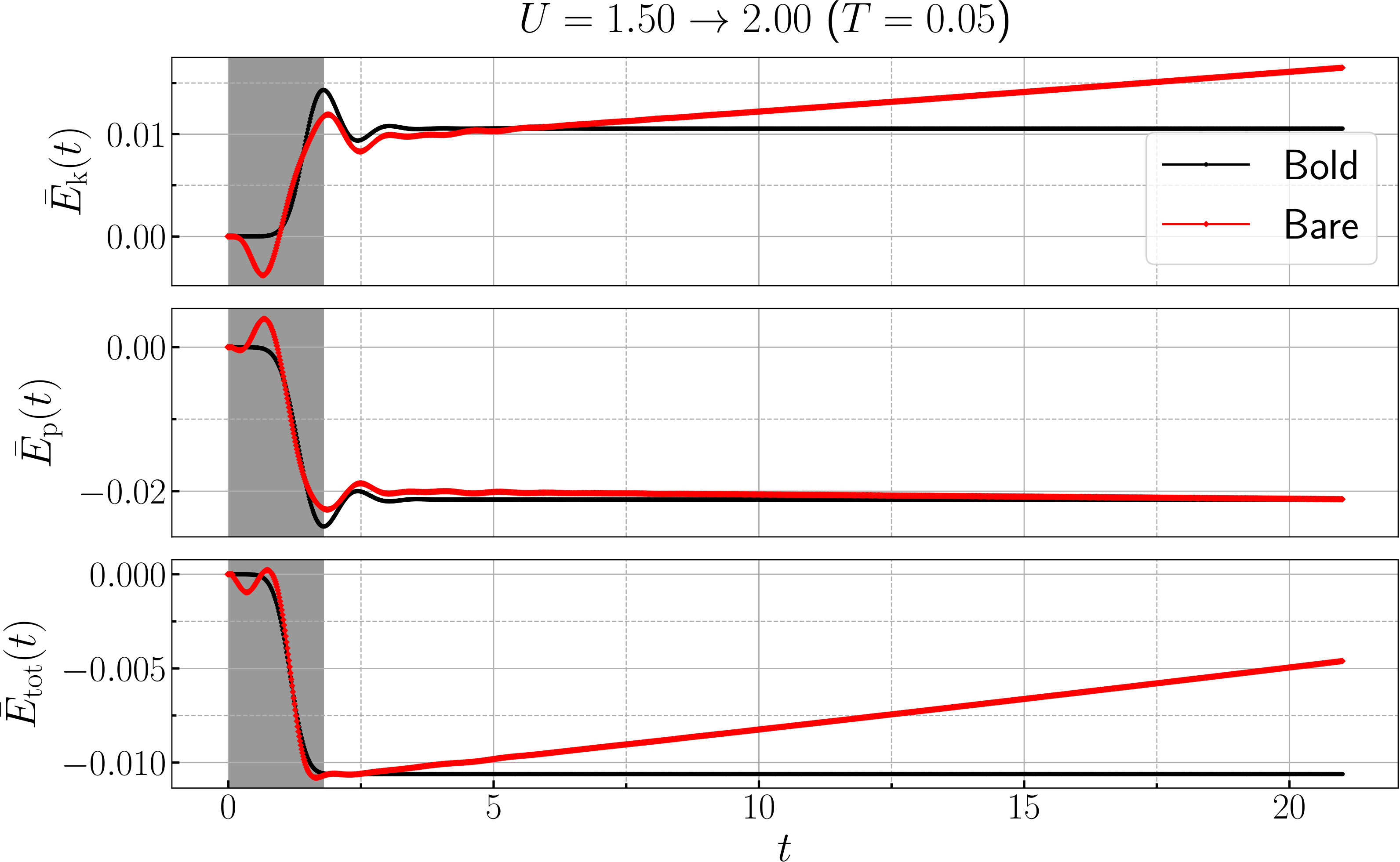}
      \caption{Energies as a function of time for an interaction ramp from $U=1.5$ to $2.0$ and initial $T=0.05$. Upper panel: Change in the kinetic energy $\bar{E}_k(t)=E_k(t)-E_k(0)$. Middle panel: Change in the potential energy $\bar{E}_p(t)=E_p(t)-E_p(0)$. Lower panel: Change in the total energy $\bar{E}_{\text{tot}}(t)=\bar{E}_k(t)+\bar{E}_p(t)$. The black curves show the results for bold IPT, which conserves energy after the ramp, and the red curves the results for bare IPT. A time step $\mathrm{d}t=0.015$ is used on the real axis, and 1200 imaginary time points on the Matsubara axis to ensure the stability of the solution at longer times. The shaded area indicates the duration of the interaction ramp.}
  \label{fig:energy_bold_bare_comparison}
\end{figure}

\section{Results}
\label{sec:results}

\subsection{General remarks}

We compute the longitudinal optical conductivity for the weakly interacting half-filled one-band Hubbard model~Eq.~\eqref{eq:Hubbard_model_intro} in dimension $D=1$ using the DMFT Green's functions obtained with non-equilibrium IPT.  We restrict the calculations to $U\lesssim \text{bandwidth}/2$, since this is the regime of parameters where our post-processing method can be expected to give sensible results.\cite{PhysRevB.103.104415_Simard_pi_ton} As discussed in Sec.~\ref{subsubsec:IPT}, there are two alternative schemes based on the self-energy $\underline{\Sigma}[\underline{\mathcal{G}}_0]$ (bare IPT) or $\underline{\Sigma}[\underline{\mathcal{G}}]$ (bold IPT). If the self-energy is expressed as a product of dressed Green's functions, energy is conserved after a $U$-quench or $U$-ramp, while the implementation with the bare Green's function $\underline{\mathcal{G}}_0$ does not conserve energy at longer times, as illustrated in Fig.~\ref{fig:energy_bold_bare_comparison} for a ramp from $U=1.5$ to $U=2.0$ with initial temperature $T=0.05$. Here, we plot the change in the kinetic energy $E_\text{k}(t)=\frac{-i}{N_{\mathbf{k}}}\sum_{\mathbf{k}}\epsilon_{\mathbf{k}}\mathcal{G}^{<}_{\mathbf{k}}(t,t)$, the potential energy $E_\text{p}(t)=\frac{-i}{2}\int_{\mathcal{C}}\mathrm{d}z \ \left[ \Sigma(t,z) \mathcal{G}(z,t)\right]^{<}(t,t)$, and the total energy $E_\text{tot}(t)=E_\text{k}(t)+E_\text{p}(t)$. Although bare IPT is more accurate than bold IPT for short times,\cite{Eckstein2010ipt} the deviations in the parameter regime considered are rather small, so that we will use the conserving bold IPT scheme in the following calculations. It is important to note that the conductivity results obtained for early times using the bare IPT differ from those obtained with bold IPT only in the amplitude of the peaks: their energy positions and trend in time remain the same. Bold IPT is used mainly because it allows us to uniquely define the temperature of the thermalized state after the ramp/quench.
 
For consistency between the DMFT and the post-processing calculations, one needs to use renormalized interactions in the latter. For a given $U$, the renormalized interaction $U_\text{ren}$ in the $\pi$-ton ladder is defined such that the divergence of the ladder contribution is shifted to the Ne\'el temperature $T_N$. We use here the renormalized $U_\text{ren}$ determined in Ref.~\onlinecite{PhysRevB.103.104415_Simard_pi_ton}, which are $U_{\text{ren}}=1.33$ for $U=2$, $U_{\text{ren}}=1.10$ for $U=1.5$ and $U_{\text{ren}}=U$ for $U=1$. Up to some rescaling, the shape of the ramp profile used for the renormalized interaction is the same as that used for the bare interaction.

\begin{figure}[b]
  \centering
    \includegraphics[width=1.0\linewidth]{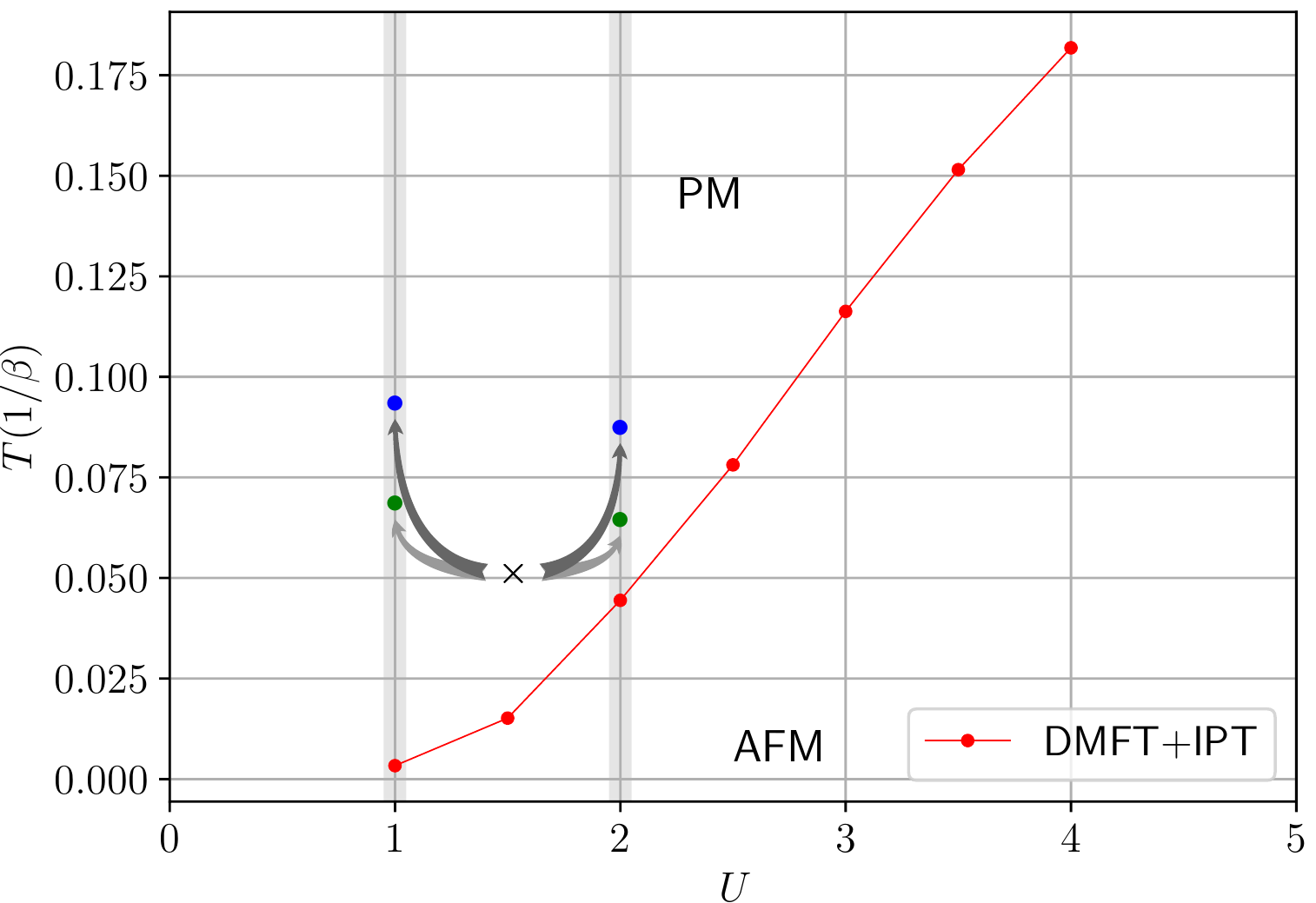}
      \caption{Sketch of the interaction ramps and quenches discussed in the present paper. The green dots (blue dots) show the temperatures of the thermalized systems after the $U$-ramps ($U$-quenches), while the red line shows the  DMFT+IPT antiferromagnetic phase boundary taken from Ref.~\onlinecite{PhysRevB.103.104415_Simard_pi_ton}. The black cross represents the initial state ($U=1.5, T=0.05$). The vertical grey lines indicate the final values of the interaction ramps/quenches.}
  \label{fig:phase_diagram_ramps}
\end{figure}

We restrict ourselves to $U\in \left[1,2\right]$, where the N\'eel temperature is not too low, because this allows us to stay in proximity to the AFM phase boundary and still have a stable time propagation with a reasonably large time step. The latter is important to access long enough times for a meaningful Fourier analysis. Moreover, within the range of bare interactions considered, the local irreducible vertices in both the charge and spin channels do not 
differ much from each other and are close to the bare interaction value.\cite{Vilk_1997} Specifically, we will consider (i) a ramp and quench up from $U=1.5$ to $U=2$ and (ii) a ramp and quench down from $U=1.5$ to $U=1$, both starting at $T=0.05$ (see black cross and arrows in Fig.~\ref{fig:phase_diagram_ramps}). Since energy is injected into the system by the ramp or quench, the temperature $T_\text{therm}$ after thermalization will be higher than in the initial state. To determine $T_\text{therm}$, we compute the total energy $E_+=E_\text{tot}(\tau_+)$, which due to the bold IPT solver is conserved after the ramp ($t\ge \tau_+$), and search for the temperature of the equilibrium system with the post-ramp $U$ and $E_\text{tot}=E_+$. For the ramp (quench) up, this calculation yields $T_\text{therm}=0.0616$ ($T_\text{therm}=0.0852$) and for the ramp (quench) down $T_\text{therm}=0.0664$ ($T_\text{therm}=0.0909$). In Fig.~\ref{fig:phase_diagram_ramps} we sketch the two ramps (quenches) together with the AFM phase boundary from Ref.~\onlinecite{PhysRevB.103.104415_Simard_pi_ton} in the plane of $U$ and $T$. The initial state of the system is indicated by the black cross, and the state of the final thermalized system is shown by the green dots (blue dots) in the case of the ramp (quench).

In the quench case, the energy injected into the system at time $t=0_+$ is given by $\Delta E_\text{tot} = \Delta U d(t=0)$, with $d(t=0)$ the double occupation in the initial state, while for an adiabatically slow ramp, the temperature of the system would be determined by the conservation of entropy. The constant entropy contours have a negative slope in the $T$-$U$ region considered in this study.\cite{PhysRevLett.95.056401} This negative slope explains why the heating effect for down ramps/quenches is stronger than for up ramps/quenches.

\subsection{Single-particle spectrum}

\begin{figure}[b]
  \centering
    \includegraphics[width=1.0\linewidth]{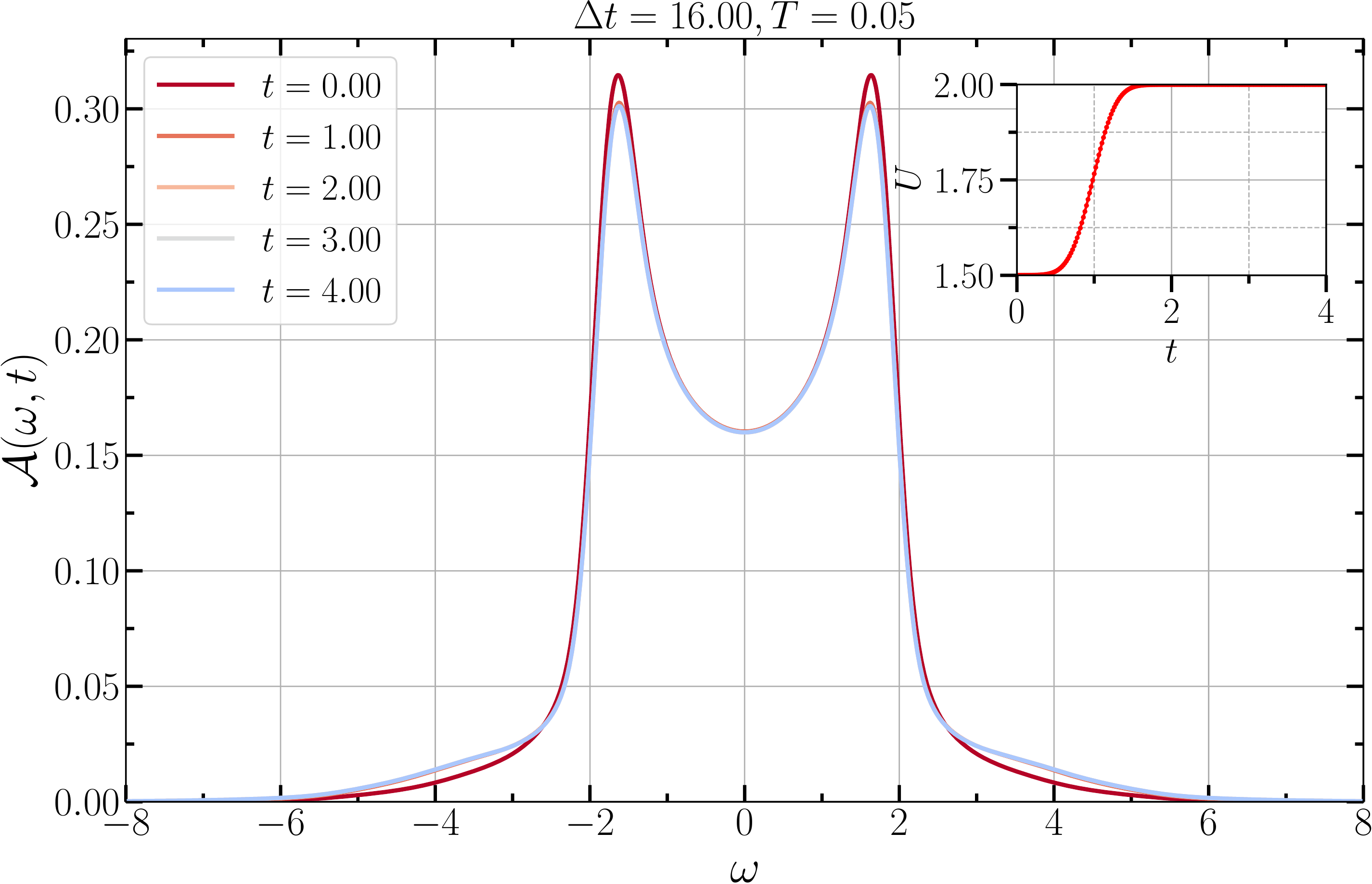}
      \caption{Illustration of the time-dependent single-particle spectral function during and after the up ramp. The inset shows the profile of the interaction ramp. The temperature of the initial state is $T=0.05$ and the Fourier window is $\Delta t=16$.
      }
  \label{fig:one_body_spectral_weight}
\end{figure}

The correction to the optical conductivity \eqref{eq:sus:non_eq_sl_KB_vertx_corr} depends on the single-particle propagator $\mathcal{G}$, which enters the calculation of the RPA-type ladder. Therefore, one can suspect that the properties of the spectral function will leave some traces in the conductivity. For that reason, we show in Fig.~\ref{fig:one_body_spectral_weight} an example of the local single-particle spectral function 
\begin{equation}
\label{eq_spectral}
\mathcal{A}(\omega,t)\equiv \mathcal{A}^R(\omega,t)=-\frac{1}{\pi}\text{Im}\int_t^{t+\Delta t} dt' \mathcal{G}_\text{loc}^R(t,t')e^{i\omega(t-t')}
\end{equation}
for various times during and after the interaction ramp from $U=1.5$ to $2$ (see inset). 

Figure~\ref{fig:one_body_spectral_weight} shows that the van Hove singularities at $\omega=\pm 2$ in the noninteracting DoS are broadened by the interactions, and shifted to $\omega \simeq \pm 1.6$. There is also a shift of spectral weight to higher energies (e.~g. $2.2\lesssim \omega \lesssim 6$) with increasing $U$. These features can be interpreted as satellites of the main peaks  which are split off by an energy $\sim U$. The upper satellite corresponds to electron insertion plus creation of a short-lived ``doublon-holon" pair. The DoS thermalizes rapidly so that no significant evolution in the spectral weight can be observed after $t=2$, and the spectra coincide with those of the thermalized system. 

\begin{figure}[b]
  \centering
    \includegraphics[width=1.0\linewidth]{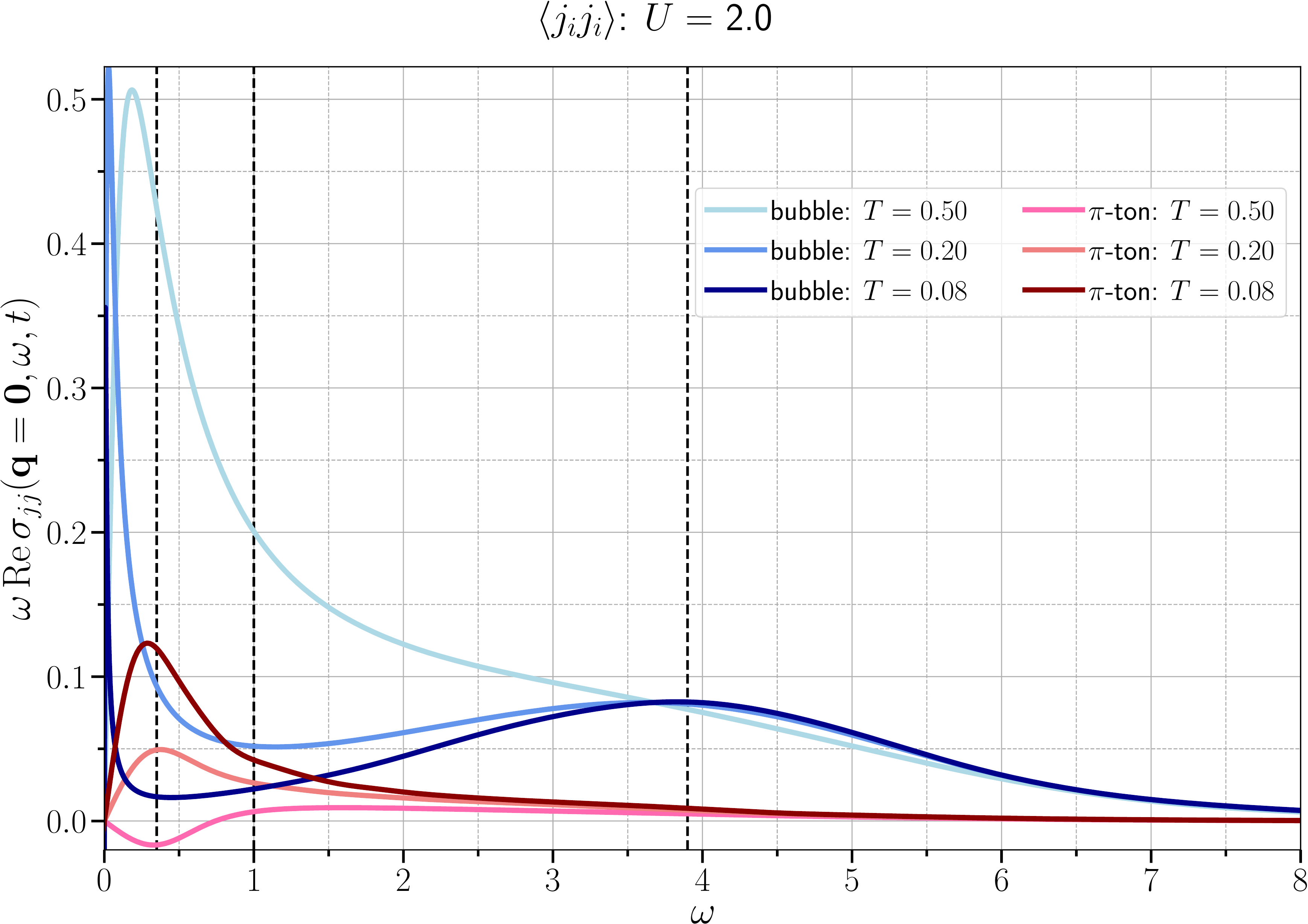}
      \caption{Longitudinal optical conductivities for both the bubble (blue shades) and the $\pi$-ton (red shades) contributions in equilibrium for different temperatures and $U=2$. (The $U=1$ results show the same qualitative trend, although the temperature scales are lower.) The vertical dotted lines indicate the energies for which we compute the time evolution of the spectral weight.
      }
  \label{fig:equilibrium_T_ramp}
\end{figure}

\subsection{Optical conductivity}
\label{subsec:optical_conductivity}

\subsubsection{Equilibrium spectra}

Prior to analyzing the non-equilibrium evolution of the optical conductivity $\text{Re}\sigma_{jj}({\bf q=0},\omega)=\text{Im}\chi_{j_ij_i{\bf q=0}}(\omega)/\omega$ we show in Fig.~\ref{fig:equilibrium_T_ramp} the equilibrium result at $U=2$ for different temperatures. Blue lines show the bubble contribution Eq.~\eqref{eq:sus:non_interacting_bubble} and red lines the $\pi$-ton contribution Eq.~\eqref{eq:sus:non_eq_sl_KB_vertx_corr}. Since we multiply the conductivity by $\omega$, the Drude peak is cut off and the figure emphasizes the spectral weight distribution at higher energies. With increasing temperature, the Drude peak shrinks but broadens, which leads to a significant increase in the bubble contribution at low, but not too small energies, as seen in Fig.~\ref{fig:equilibrium_T_ramp}. The broad peak in the bubble contribution near $\omega=3.9$ can be associated with excitations between the peaks in the DoS (Fig.~\ref{fig:one_body_spectral_weight}).

We note that in order to resolve the low-frequency behavior, a very large time-window is needed. The bubble results in Fig.~\ref{fig:equilibrium_T_ramp} were obtained by extrapolating the calculated equilibrium data to long times with an exponential fit. In the case of the $\pi$-ton contribution, oscillations persist to much longer times, so that we show the Fourier transformation on the calculated time window $\Delta t=17$, which implies some smearing of the low-frequency features. 

The spectrum obtained from the $\pi$-ton contribution shows a nontrivial temperature dependence at low frequencies, but the main characteristic feature is a peak near $\omega=0.35$, which grows as one approaches the AFM phase boundary at low temperatures. At high temperatures, this peak switches from positive to negative, which implies that the $\pi$-ton narrows (broadens) the Drude feature at high (low) $T$. 

Figure~\ref{fig:equilibrium_T_ramp} shows that, near the AFM boundary, the equilibrium optical conductivity (including vertex correction) is composed of (i) a low-energy Drude peak, which depending on the temperature range can be enhanced or narrowed by the $\pi$-ton type vertex correction, and (ii) a broad high-energy hump near $\omega=3.9$, originating mainly from the bubble diagram and related to peaks in the single-particle DoS. These results are qualitatively and quantitatively consistent with the conclusions reached in Ref.~\onlinecite{PhysRevB.103.104415_Simard_pi_ton} based on less reliable Maximum Entropy analytical continuation of imaginary time data. The general features and trends are also consistent with the $\pi$-ton related modifications of the conductivity observed in Refs.~\onlinecite{kauch_pitons_2019,worm2020broadening}, which used more systematic methods involving parquet equations, and a semi-analytical RPA evaluation of the $\pi$-ton, respectively. Similar observations related to the longitudinal conductivity were also reported in Ref.~\onlinecite{Bergeron_2011_optical_cond}.

\begin{figure}[t]
  \centering
    \includegraphics[width=\linewidth]{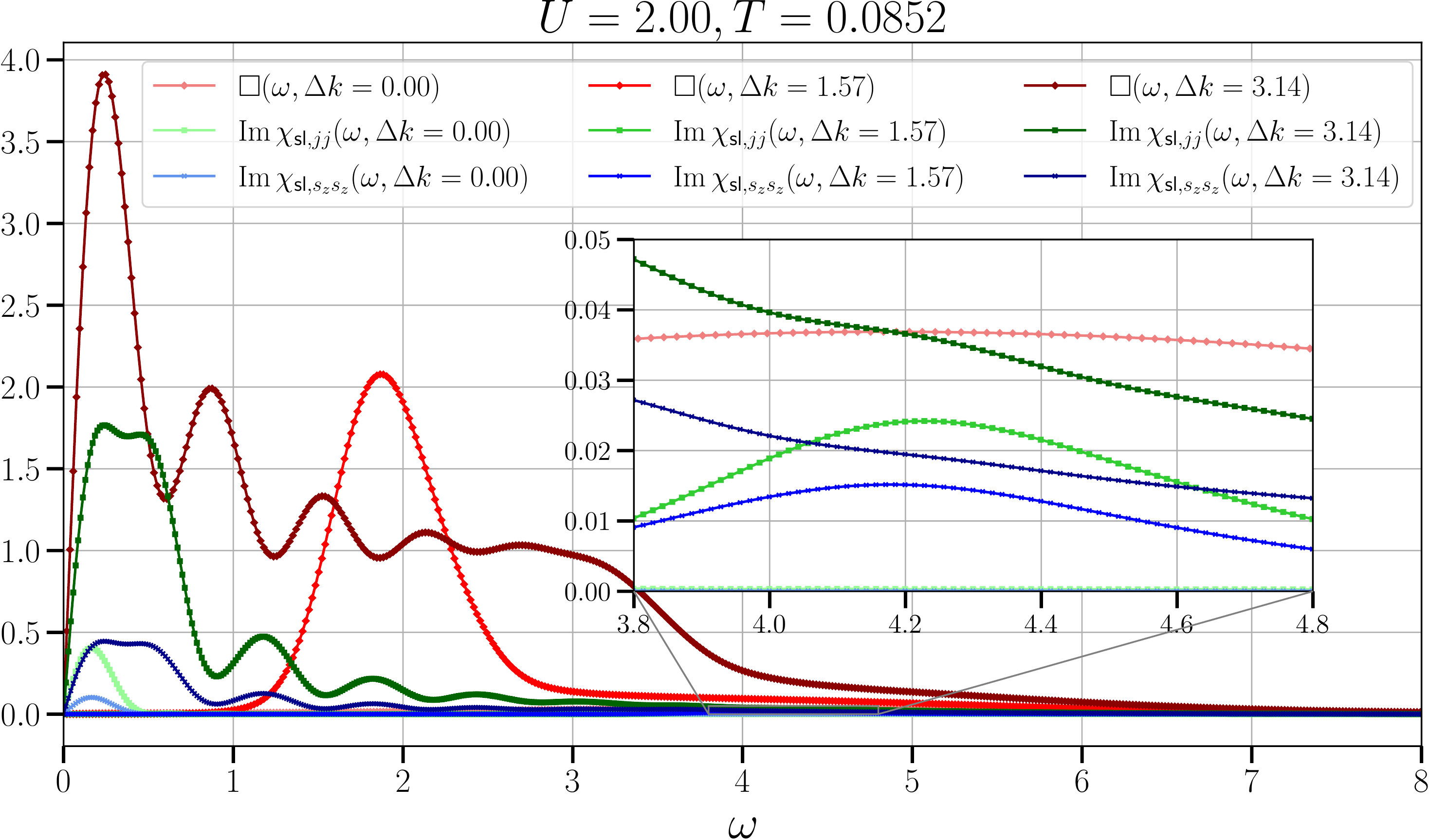}
      \caption{Analysis of different contributions to the $\pi$-ton spectrum. The different shades of red show the momentum dependence of the single-ladder vertex spectrum. Light, intermediate and dark colored lines show the contributions from momentum tuples with $\Delta\mathbf{k}=|\tilde{\mathbf{k}}-\bar{\mathbf{k}}| = 0$, $\frac{\pi}{2}$, and $\pi$, respectively. The shades of green show the momentum dependence of Im$\chi_{\text{sl},j_ij_i}$, whereas the different shades of blue show that of Im$\chi_{\text{sl},s_zs_z}$ (vertex correction only).}
  \label{fig:k_decomposition_various_components}
\end{figure}

To better understand the characteristic energy scales of the $\pi$-ton contribution to the conductivity, we show in Fig.~\ref{fig:k_decomposition_various_components} the reducible single-ladder vertex ``$\square$" appearing in Eq.~\eqref{eq:sus:vertical_ladder_sin_vie2_explicited} for the three momentum differences $|\tilde{\mathbf{k}}-\bar{\mathbf{k}}|\in \{0,\frac{\pi}{2},\pi\}$ (red shaded lines). In addition, we plot both the imaginary parts of the current and spin susceptibilities for the indicated momenta to illustrate the effect of multiplying the four Green's functions in Eq.~\eqref{eq:sus:non_eq_sl_KB_vertx_corr} and adding velocity factors at the vertices. The plotted Im$\chi_{\text{sl},j_ij_i}$ (green shades) and Im$\chi_{\text{sl},s_zs_z}$ (blue shades) represent the $\pi$-ton contribution \eqref{eq:sus:non_eq_sl_KB_vertx_corr} associated with the $(\tilde{\mathbf{k}},\bar{\mathbf{k}})$-tuples whose difference corresponds to $\Delta\mathbf{k}=|\tilde{\mathbf{k}}-\bar{\mathbf{k}}| \in \{0,\frac{\pi}{2},\pi\}$, namely $\frac{1}{N_{\mathbf{k}}}\sum_{|\tilde{\mathbf{k}}-\bar{\mathbf{k}}|=\Delta\mathbf{k}}\underline{\chi}_{\text{sl}}(\mathbf{\tilde{k}},\mathbf{\bar{k}},\mathbf{q}=\mathbf{0};\omega)$. The spin-spin single-ladder vertex correction $\underline{\chi}_{\text{sl},s_zs_z\mathbf{q}}$ is equal to Eq.~\eqref{eq:sus:non_eq_sl_KB_vertx_corr} with a global factor of $-1$ and without the velocity factors.

In the momentum-dependence of $\square$, one notices a prominent peak appearing around $\omega\simeq2$, which originates from $\mathbf{k}=\frac{\pi}{2}$. This feature is suppressed once the four Green's functions are multiplied to the ladder, as defined in Eq.~\eqref{eq:sus:non_eq_sl_KB_vertx_corr}, independent of the presence or absence of velocity factors. On the contrary, the tiny $\mathbf{k}=\mathbf{0}$ contribution to the ladder contribution gets enhanced by the multiplication with these Green's function, especially for Im$\chi_{\text{sl},jj}$. However, as the name suggests, the by far dominant contribution to the $\pi$-ton comes from $|\tilde{\mathbf{k}}-\bar{\mathbf{k}}|=\pi$.
In the single-band nearest-neighbor Hubbard model, the Fermi momenta are $\mathbf{k}_F=\pm \frac{\pi}{2}$. These coincide with the largest values of the velocities and are separated by a momentum shift $\pi$, partly explaining why Im$\chi_{\text{sl},j_ij_i}$ is larger than Im$\chi_{\text{sl},s_zs_z}$.

Note that in $\operatorname{Im}\chi_{\text{sl},j_ij_i}(\omega,\Delta\mathbf{k}=\pi)$ (dark green spectrum), a hump appears in Fig.~\ref{fig:k_decomposition_various_components} near $\omega\simeq 1$. In the following subsection, we will thus investigate the time traces of the conductivity at $\omega=0.35$, $1.0$ and $3.9$ (see black dashed lines in Fig.~\ref{fig:equilibrium_T_ramp}).

\subsubsection{Non-equilibrium evolution}

\begin{figure}[ht]
  \centering
    \includegraphics[width=1.0\linewidth]{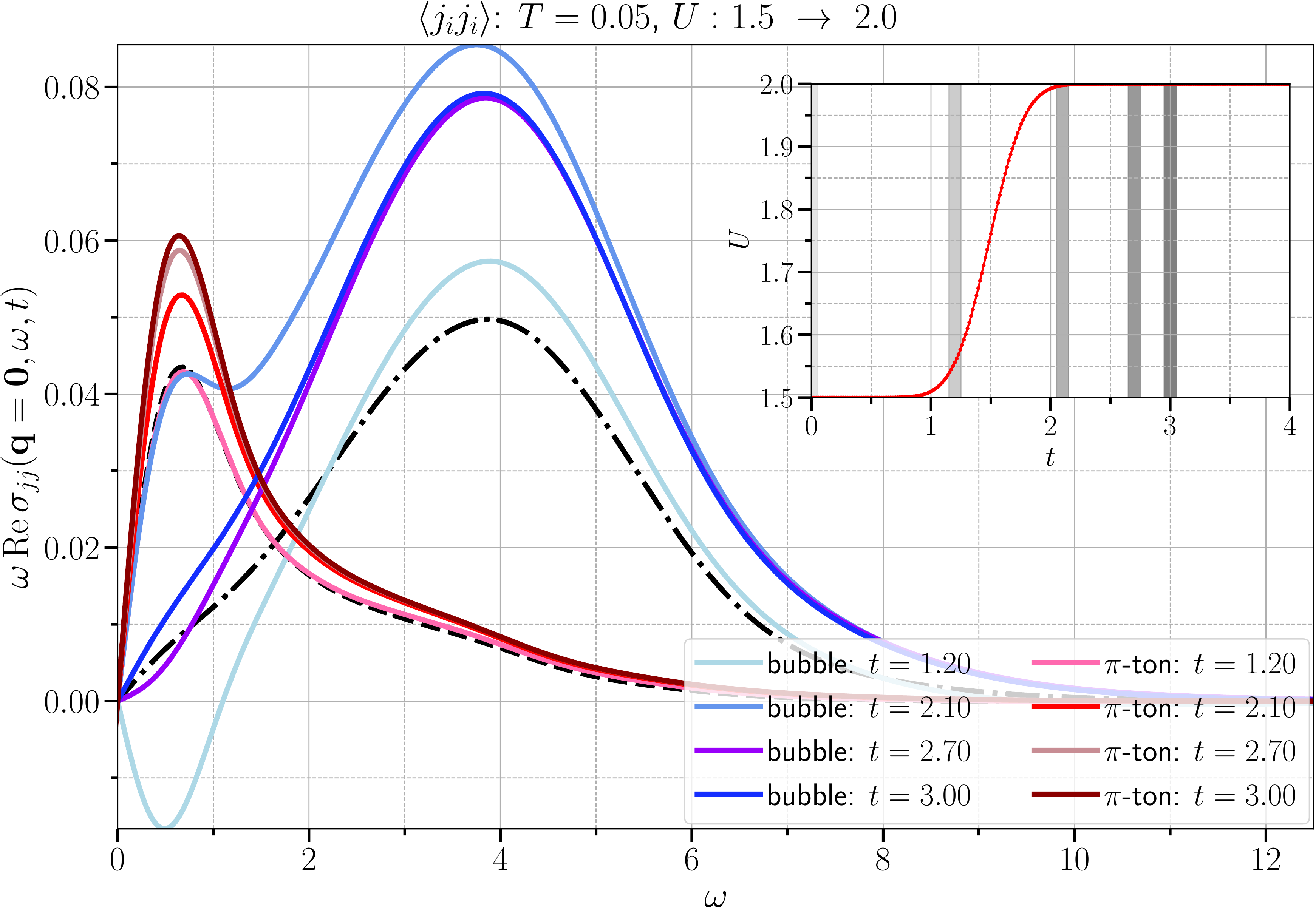}\vspace{2mm}
        \includegraphics[width=1.0\linewidth]{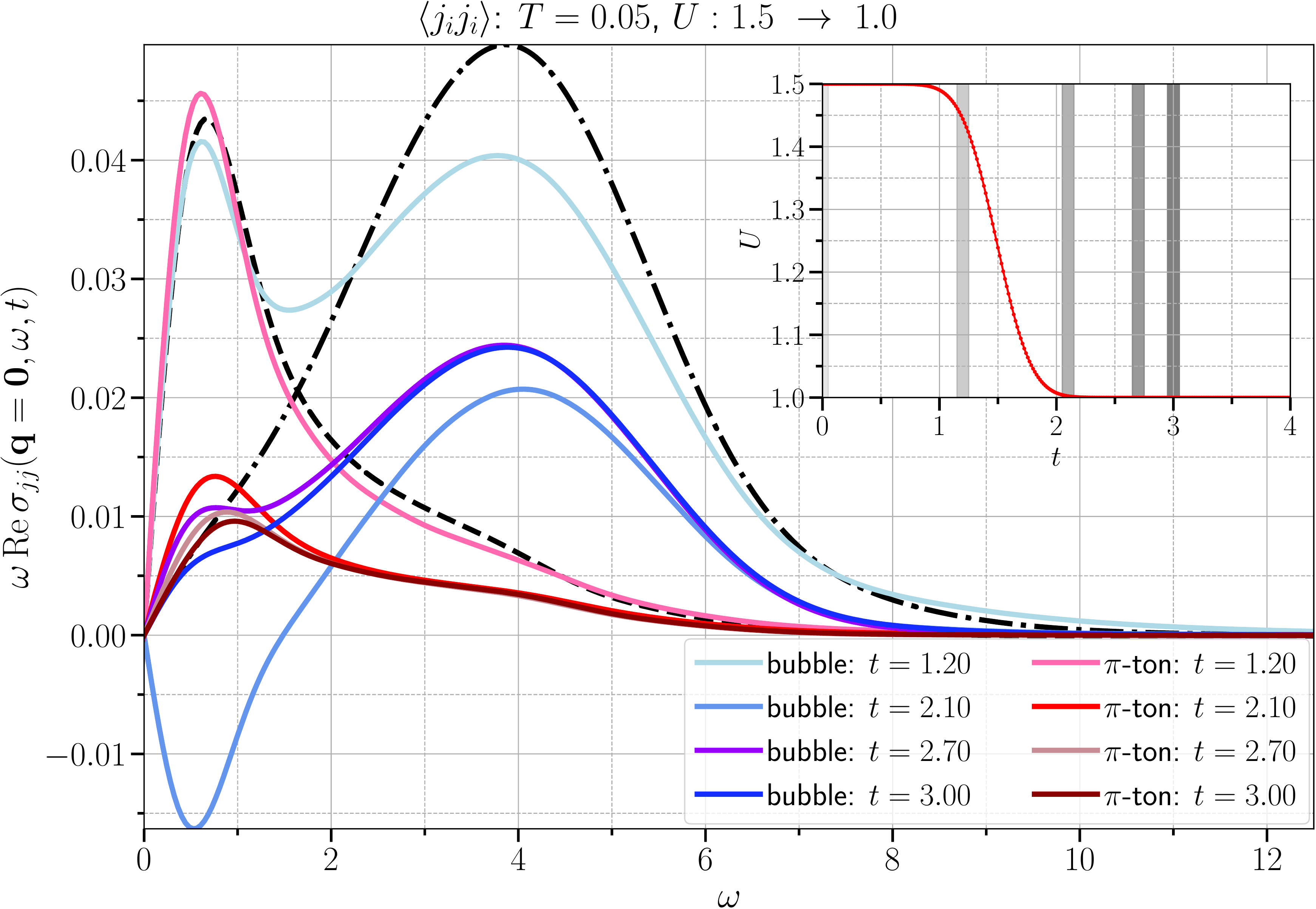}
      \caption{Real time snapshots of the bubble contribution to the optical conductivity (blue) and the RPA $\pi$-ton vertex correction (red) during and after the up ramp (top panel) and down ramp (bottom panel). The ramp shapes are shown by the inset, with grey lines indicating the measurement times. The dotted-dashed (dashed) black line shows the bubble ($\pi$-ton) contribution in the initial equilibrium state. A Fourier window $\Delta t=7$ is used to compute the equilibrium and time-dependent spectra.
     }
  \label{fig:4_panels_jj_up_1p5_2}
\end{figure}

We next investigate how the bubble and $\pi$-ton contributions to the conductivity evolve with the interaction ramps and quenches in the vicinity of the AFM phase boundary (see Fig.~\ref{fig:phase_diagram_ramps}). In Fig.~\ref{fig:4_panels_jj_up_1p5_2}, the bubble contribution to the optical conductivity is plotted with blue-shaded lines and the $\pi$-ton correction with red-shaded lines. For comparison, dotted-dashed and dashed black lines indicate, respectively, the bubble and $\pi$-ton spectra in the initial equilibrium state ($U=1.5$, $T=0.05$). Again, we plot $\omega \text{Re}\sigma_{jj}(\omega,{\bf q=0})=\text{Im}\chi_{jj{\bf q=0}}(\omega)$ so that the Drude peak is cut off at low frequencies. In both panels of Fig.~\ref{fig:4_panels_jj_up_1p5_2}, a time window $\Delta t=7$ is used for the Fourier transformation.

\begin{figure*}[ht]
  \centering
    \includegraphics[width=1.0\linewidth]{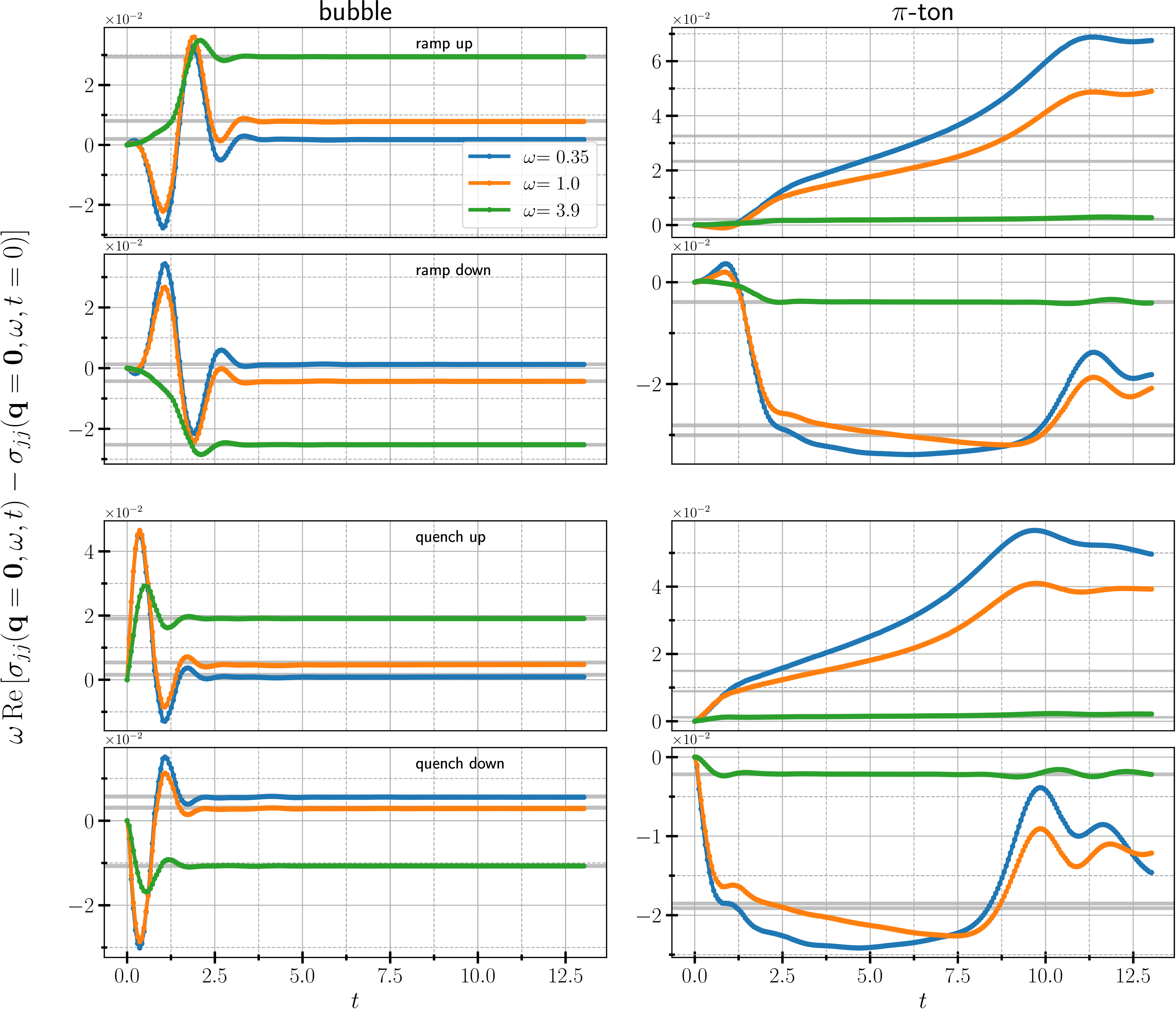}
      \caption{
      Time-dependent change in the bubble (left column) and $\pi$-ton (right panel) contributions to $\omega \text{Re}\sigma_{jj}({\bf q=0},\omega,t)$ at $\omega=0.35$ (blue), $1.0$ (orange) and $3.9$ (green). The two upper rows of panels show the results for the ramps: upper (lower) panel for the ramp up (down). Likewise, the two lower rows of panels show the results for the quenches. Horizontal grey lines indicate the values reached in the thermalized state after the ramp, with the thickness of the lines approximately representing the numerical uncertainty in determining these thermalized values.  For a better visualization of the change in conductivity, we subtract the values at $t=0$. Just like for Fig.~\ref{fig:4_panels_jj_up_1p5_2}, a time window $\Delta t = 7$ was used at all times $t$.
      }
  \label{fig:w_cut_2_down_1p5_1}
\end{figure*}

In the top panel of Fig.~\ref{fig:4_panels_jj_up_1p5_2}, we show the spectra measured at different times during and after the up ramp. The first time (light grey shade) is close to the start of the interaction ramp, while the remaining curves (darker grey shades) illustrate the evolution after the ramp. The high-energy feature in the conductivity which is associated with excitations between the van Hove singularities in the DoS (Fig.~\ref{fig:one_body_spectral_weight}) and is primarily due to the bubble contribution, shows a rapid relaxation after the ramp -- the two latest curves overlap at that energy. On the other hand, the prominent $\pi$-ton feature near $\omega\approx 0.35$ appears to relax more slowly since the two latest curves at that energy do not overlap. Also, as will become clearer in Fig.~\ref{fig:w_cut_2_down_1p5_1}, the $\pi$-ton feature does not exhibit the oscillations that appear at short times in the low-energy bubble contribution. In Fig.~\ref{fig:4_panels_jj_up_1p5_2}, those oscillations in the bubble contribution are particularly strong at earlier times where the spectra at $\omega\simeq 0.35$ change sign.

The analogous plot for the down ramp is shown in the bottom panel of Fig.~\ref{fig:4_panels_jj_up_1p5_2}. 
Regarding the bubble contribution, it reveals a qualitatively similar relaxation behavior, with oscillations in the Drude component at early times, and a more damped  relaxation of the features at higher energies. In the case of the $\pi$-ton contribution, the down ramp displays a melting of the $\pi$-ton feature at $\omega\simeq 0.35$, which appears to happen at a faster rate than that at which it builds up when ramping up the interaction. The $\pi$-ton spectral peak also shifts in energy while melting down -- more so than in the up quench.

For a more detailed analysis of the relaxation behavior, we consider in Fig.~\ref{fig:w_cut_2_down_1p5_1} the evolution of the spectral weight at the three characteristic energies $\omega=0.35$ (Drude feature), $\omega=1$ (intermediate-energy feature) and $\omega=3.9$ (high-energy feature). We first consider the bubble contribution, plotted in the left panels of Fig.~\ref{fig:w_cut_2_down_1p5_1} for the ramp up, ramp down, quench up and quench down (from top to bottom). For a better visualization, we plot the changes in the spectral weight: $\omega \text{Re}[\sigma_{jj}({\bf q=0},\omega,t)-\sigma_{jj}({\bf q=0},\omega,t=0)]$. To illustrate the thermalization dynamics, we furthermore indicate by grey horizontal lines the values reached in the thermalized state (obtained by calculating the total energy after the ramp/quench, see green and blue dots in Fig.~\ref{fig:phase_diagram_ramps}). The figure clearly reveals a single large amplitude oscillation in the Drude feature during (after) the ramps (quenches), and similar, but more strongly damped oscillations in the higher energy cuts. It also shows that after this initial oscillation, the bubble contribution to the conductivity rapidly relaxes to the thermalized result at all three energies.

The thermalized values of the conductivity for the ramp (quench) up are larger than in the initial state because correlation effects shift spectral weight to higher energies. Interestingly, though, the initial response of the Drude feature to the ramp goes in the opposite direction. For example, in the early stages of the ramp up, the weight at $\omega=0.35$ decreases substantially (while a large transient increase is found for the ramp down). However, in the case of the quenches this short-time behavior can be qualitatively different, and thus appears to be related to the details of the ramp spectrum. In the case of the down ramp or quench, we furthermore notice that the relative change of the Drude weight has the opposite sign from that expected due to correlation-induced broadening -- presumably this is because of the strong heating effect. 

\begin{figure}[ht!]
  \centering
    \includegraphics[width=1.0\linewidth]{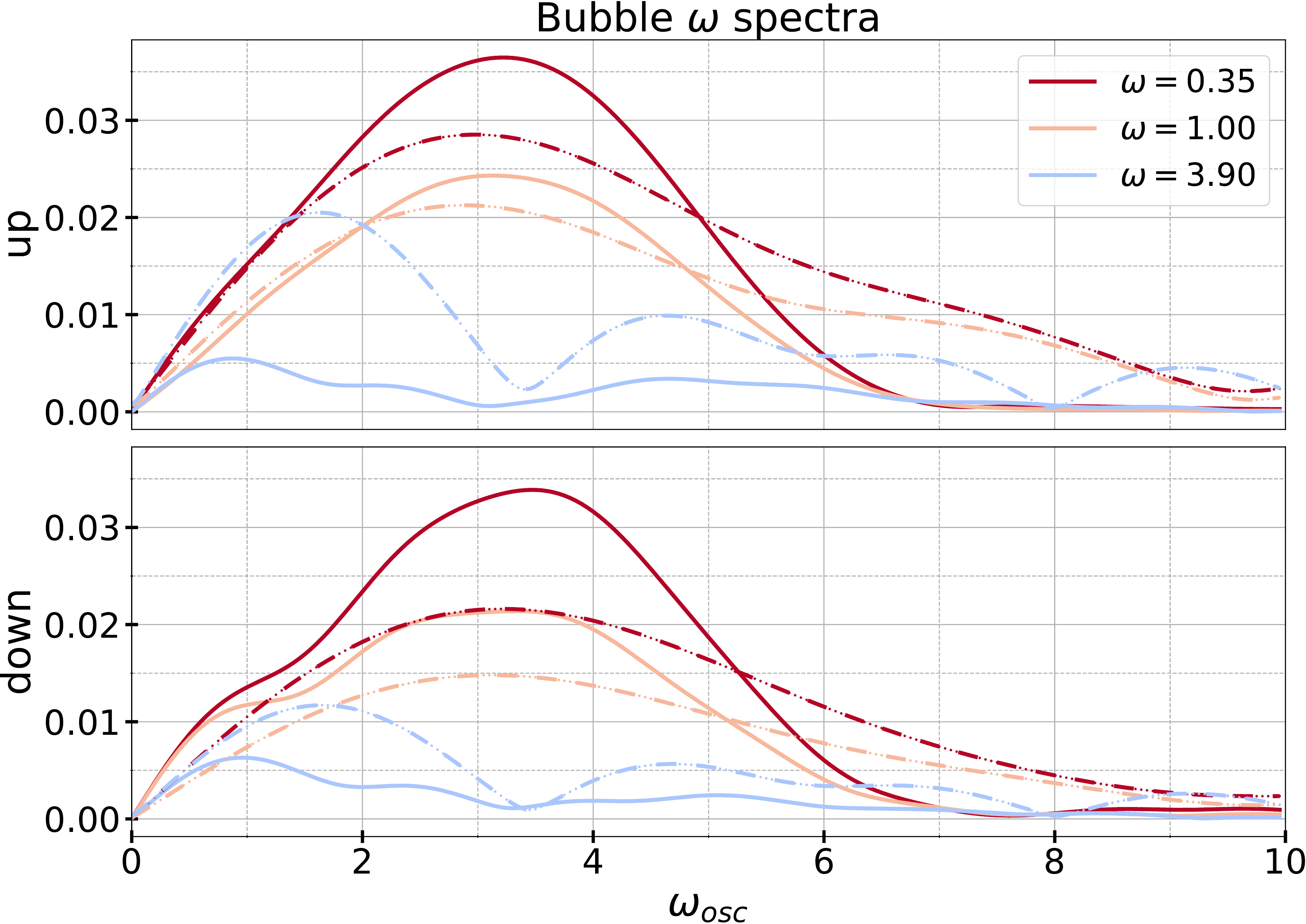}
      \caption{
      Spectral decompositions of the time traces (see Fig.~\ref{fig:w_cut_2_down_1p5_1}) of the bubble contribution $\omega \text{Re}\sigma_{jj}({\bf q=0},\omega,t)$ at the indicated energies. The plotted lines show the norms of the Fourier transformations, after subtracting a background proportional to the ramp shape (which is described by the error function). In the case of the quenches, the mean was subtracted. Dashed (solid) lines are for interaction ramps (quenches). 
      }
  \label{fig:w_cut_spectral_FFT}
\end{figure}

The ramp and quench induced changes in the $\pi$-ton contribution to the conductivity at $\omega=0.35$, $1$ and $3.9$ are plotted as a function of time in the right panels of Fig.~\ref{fig:w_cut_2_down_1p5_1}. The green lines, corresponding to $\omega=3.9$, indicate that the high-energy structures of the $\pi$-ton approach the thermal value quickly after the ramp or quench, on a timescale comparable to the bubble contribution. This is different for the intermediate-energy and Drude features. The latter exhibit a delayed thermalization, overshooting the thermal reference values by a significant amount, especially when quenching/ramping the interaction up. For some of these traces the initial response of the $\pi$-ton to the ramp is qualitatively similar to that of the bubble contribution, in the sense that the transient change of the spectral weight goes in the opposite direction from the modification expected in the thermalized state, but the effect is much less pronounced than for the bubble.

The substantial overshooting of the thermal reference values could be a manifestation of prethermalization behavior. The optical conductivity is related to the kinetic energy via a sum rule, and the energy distribution function is known to exhibit a prethermalization plateau at low energies after quenches in the considered interaction regime.\cite{Moeckel_2008,Eckstein_2009}  
In the following section, we will investigate the occupation and non-equilibrium distribution functions to demonstrate that in contrast to the spectral function, which thermalizes fast (Fig.~\ref{fig:one_body_spectral_weight}), the occupation remains nonthermal for a long time. The $\pi$-ton correction to the conductivity depends strongly on the occupied density of states near the Fermi level, due to a combination of the facts that the velocities are largest in absolute values near the Fermi level and that the poles of the single-ladder vertex dominate for momentum differences equal to $\mathbf{k}_{\pi}$. We will thus look at the distribution for ${\bf k}\approx \frac{\pi}{2}$ and discuss its repercussions in Sec.~\ref{sec:discussion}.

We also notice significant modifications in the $\pi$-ton contribution after $t\approx 10$, especially in the low- and intermedite-energy cuts. This is partly related to the energy shifts in the $\pi$-ton feature which at early times are clearly seen in the bottom panel of Fig.~\ref{fig:4_panels_jj_up_1p5_2}. We will discuss other possible explanations for this behavior below in Sec.~\ref{sec:discussion}. We note that the time traces shown in Fig.~\ref{fig:w_cut_2_down_1p5_1} are converged with respect to the time step used in the real-time propagation. Changing the step size within the range $0.01\leq \mathrm{d}t \leq 0.03$ does not affect our observations.


\section{Discussion}
\label{sec:discussion}

In this section, we analyze the results presented in Sec.~\ref{sec:results} and extract more information on characteristic oscillation frequencies and the thermalization behavior.

Notably, we have identified a qualitatively different time evolution of the bubble contribution and RPA $\pi$-ton type vertex correction to the optical conductivity after an interaction ramp or quench in the vicinity of the AFM phase boundary at weak coupling. To further investigate the dynamics of these features, we first extract the dominant oscillation frequencies $\omega_\text{osc}$ in the bubble signals in Fig.~\ref{fig:w_cut_2_down_1p5_1} by performing Fourier transformations on the time traces shown in the left panels. Here, in order to remove spurious spectral weight coming from the ramp, we subtract from the signal a smooth background proportional to the ramp profile. The norms of the Fourier transforms of the bubble contribution are shown in Fig.~\ref{fig:w_cut_spectral_FFT}, where solid lines indicate the spectra for the ramps and dashed lines those for the quenches. The upper (lower) panels show the oscillation spectra for the up (down) ramp/quench at $\omega=0.35$ (red), $1.0$ (orange) and $3.9$ (blue). The spectra in Fig.~\ref{fig:w_cut_spectral_FFT} reveal the main oscillation frequencies induced by the perturbations. 

For $\omega=0.35$ (Drude component) and $1.0$ (intermediate-energy peak), independent of the direction of the ramp/quench, the oscillations yield a single peak centered at $\omega_\text{osc}\approx 3.2$, which roughly matches the energy separation between the peaks in the DoS (see Fig.~\ref{fig:one_body_spectral_weight}). This peak is very broad, because the corresponding time traces are strongly damped after the first oscillation. As one can already deduce from the time traces (Fig.~\ref{fig:w_cut_2_down_1p5_1}), the amplitude of the oscillations is a bit smaller for the quench (dashed lines) than for the ramps (solid lines). At $\omega=3.9$ the damping is even stronger and the thermalization faster. While the oscillation frequency in the time traces seems to be the same as for the lower energy cuts, the subtraction of the smooth background in the form of the ramp shape results in a spectrum consisting of two frequency humps centred around $\omega\simeq 1.5$ and $\omega\simeq 5$. These may correspond to fluctuations between the DoS peaks and the side-bands visible in Fig.~\ref{fig:one_body_spectral_weight}. Also, for $\omega=3.9$, the amplitude of the oscillations is larger for the quench than for the ramp. 

As already mentioned above, the $\pi$-ton time traces shown in Fig.~\ref{fig:w_cut_2_down_1p5_1} exhibit no pronounced oscillations, but rather a prethermalization behavior, especially at $\omega=0.35$ and $1.0$, and in the case of the up ramp/quench. The thermalization of this vertex correction to the optical conductivity occurs on timescales which are much longer than the accessible simulation times. Furthermore, the prethermalization phenomenon is whittled down when the temperature is raised (not shown).

To shed some light on the origin of the prethermalization behavior, we take a closer look at the time evolution of the different components of $\mathcal{G}$. We do so because the RPA $\pi$-ton vertex correction is built from non-equilibrium Green's functions $\mathcal{G}$ computed within DMFT. Since the spectral function, extracted from the retarded component $\mathcal{G}^R$, thermalizes fast (Fig.~\ref{fig:one_body_spectral_weight}), the trapping in a prethermalized state must be primarily due to nonthermal properties of $\mathcal{G}^<$, i.e. the corresponding spectral function (occupation function) $\mathcal{A}^<$. Here, we will investigate the non-equilibrium distribution function, which allows us to establish how fast the system reaches the (momentum independent) thermalized Fermi distribution function $n_F(\omega) = \frac{1}{e^{\beta\omega}+1}$.

The retarded spectral function reads
\begin{align}
\mathcal{A}^{R}_\mathbf{k}(t,\omega) = -\frac{1}{\pi}\operatorname{Im}\mathcal{G}^{R}_\mathbf{k}(t,\omega),
\end{align} 
and the lesser spectral function is defined as
\begin{align}
\mathcal{A}^{<}_\mathbf{k}(t,\omega) = \frac{1}{2\pi}\operatorname{Im}\mathcal{G}^{<}_\mathbf{k}(t,\omega),
\end{align} 
where we use a forward-in-time Fourier transformation as in Eq.~\ref{eq_spectral} with a large cut-off in time. Large time windows are accessible because the real-time functions $\mathcal{G}^{<,R}_{\mathbf{k}}(t,t^{\prime})$ for fixed time $t$ have tails which can be fitted by the function $\alpha e^{-t^{\prime}/\beta}\cos{(\varepsilon t^{\prime}+\delta)}$. We can therefore extrapolate these functions before the Fourier transformation. From both the lesser and retarded spectral functions, a non-equilibrium distribution function $n_\mathbf{k}$ can be computed as\cite{RevModPhys.86.779_non_eq_review}
\begin{align}
n_\mathbf{k}(t,\omega) = \frac{\mathcal{A}^<_\mathbf{k}(t,\omega)}{\mathcal{A}^R_\mathbf{k}(t,\omega)},
\end{align}
since the occupied states are given by $\text{Im}\mathcal{G}_{\mathbf{k}}^<(t,\omega)=2\pi\mathcal{A}_{\mathbf{k}}^{R}(t,\omega)n_\mathbf{k}(t,\omega)$. Because the nonthermal distribution functions are typically not of the Fermi-Dirac form, we evaluate the effective inverse temperature $\beta_\text{eff}=1/T_\text{eff}$ from the derivative of $n_\mathbf{k}(t,\omega)$ at $\omega=0$ as 
\begin{align}
\label{w_0_n_derivative_for_T}
\beta_\text{eff} \equiv -4\frac{\partial n_\mathbf{k}(t,\omega)}{\partial\omega}\bigg\rvert_{\omega=0}.
\end{align} If $n_\mathbf{k}(t,\omega)$ is of the Fermi-Dirac form $n_F(\omega)$ introduced above, Eq.~\eqref{w_0_n_derivative_for_T} yields the corresponding inverse temperature $\beta$.

\begin{figure}[ht]
  \centering
  \includegraphics[width=1.0\linewidth]{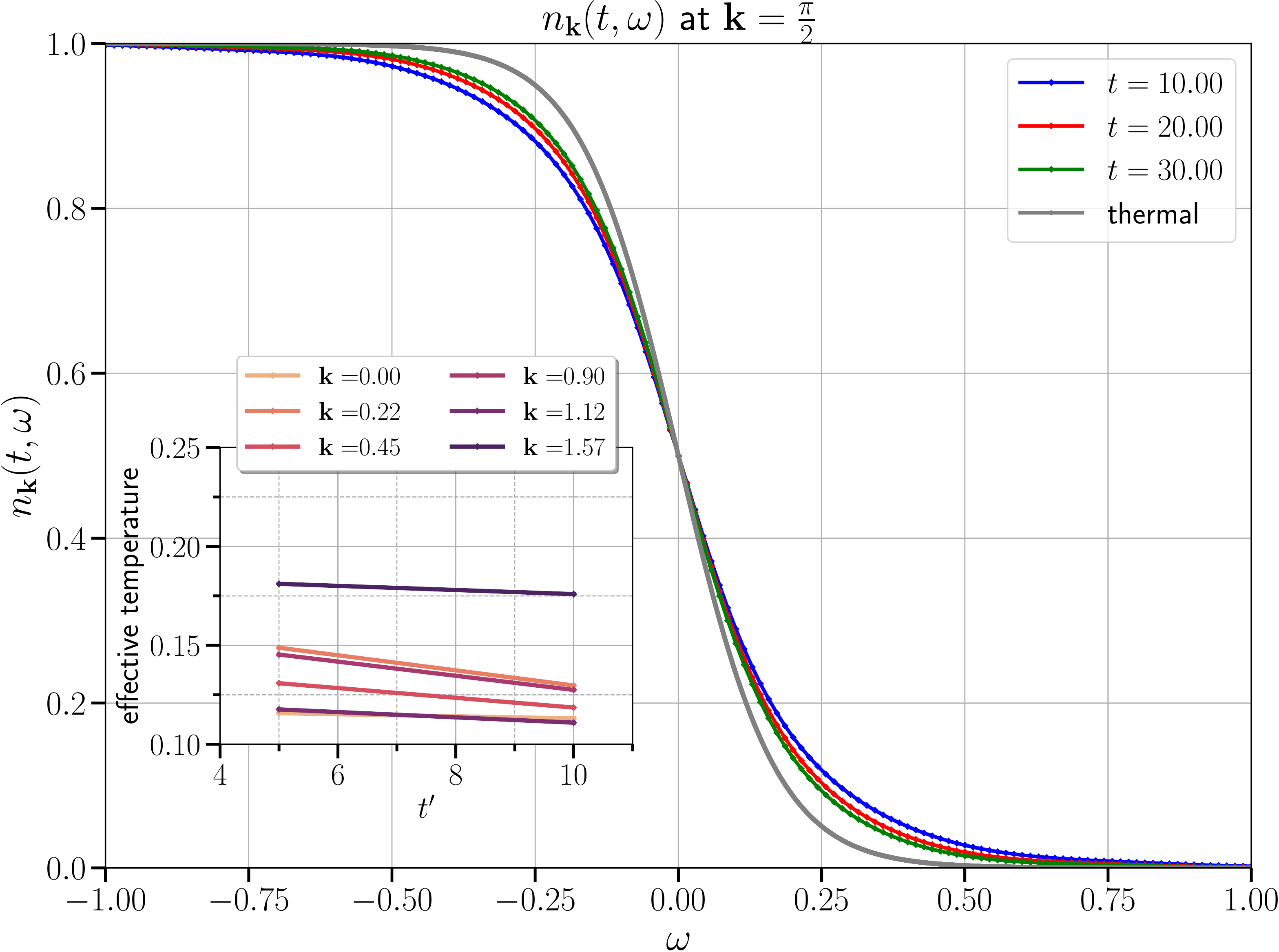}\vspace{2mm}
  \includegraphics[width=1.0\linewidth]{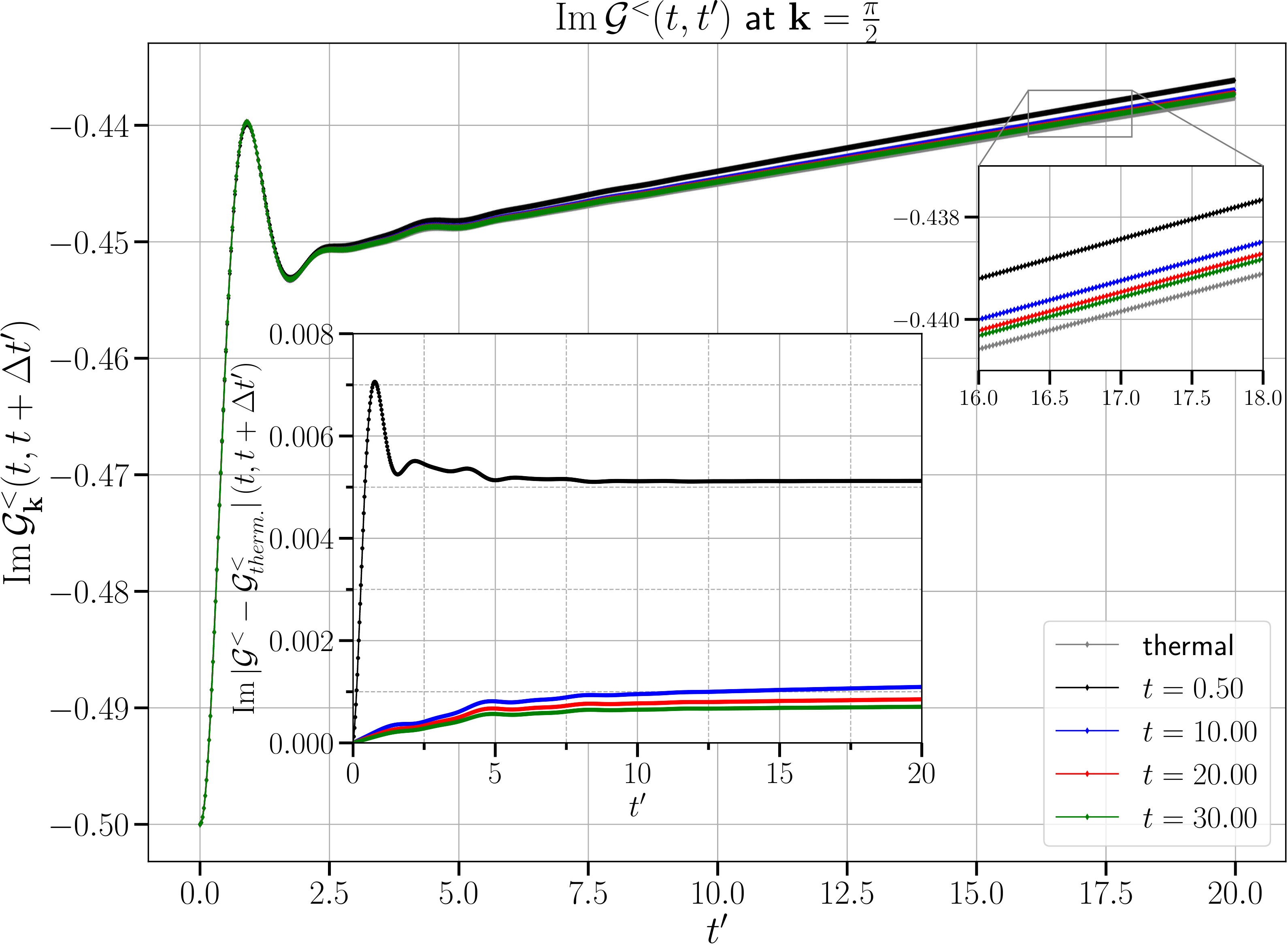}
  \caption{Top panel: In the main plot, the grey curve depicts the Fermi distribution function for the thermalized state after the quench from $U=1.5$ to $U=2$ at $T=0.05$. The other curves illustrate the non-equilibrium distributions $n_\mathbf{k}(t,\omega)$ at $\mathbf{k}=\tfrac \pi 2$ (Fermi level) and for $t=\{10,20,30\}$. An  exponential function was used to extrapolate the tails of both $\mathcal{A}^<(t,t+\Delta t^{\prime})$ and $\mathcal{A}^{R}(t,t+\Delta t^{\prime})$. In the Fourier transformation, we use a time window $\Delta t^{\prime}=4000$. In the inset plot, the time traces of the effective temperatures Eq.~\eqref{w_0_n_derivative_for_T} of the $\mathbf{k}$-dependent distribution functions are plotted. Bottom panel: In the main plot, the time evolution of the lesser component of the dressed Green's function for the Fermi momentum $\mathbf{k}=\frac{\pi}{2}$ is plotted for the same quench. 
The inset shows the difference between the time traces of the quenched system and the thermalized system ($T=0.0852$ and $U=2.0$).}
  \label{fig:fermi_dist_and_trapping_analysis_A_lesser}
\end{figure}

In the top panel of Fig.~\ref{fig:fermi_dist_and_trapping_analysis_A_lesser}, we plot the nonthermal distribution functions $n_{\mathbf{k}}(t,\omega)$ for $\mathbf{k}=\frac{\pi}{2}$ at times $t=10$, $20$ and $30$, for the quench from $U=1.5$ to $U=2$ at initial temperature $T=0.05$. We also show in grey the Fermi distribution function at the thermalized temperature $T_{\text{therm}}=0.0852$. For $\mathbf{k}=\frac{\pi}{2}$, the distributions at $t=30$ are still clearly nonthermal. Previous studies have already reported similar phenomena where an interaction quench in the weak coupling regime led to non-thermal stationary distributions at the Fermi level.\cite{Moeckel_2008,Manmana_2007,Kollath_2007} For momenta different from $\mathbf{k}=\pm\frac{\pi}{2}$, one finds a lower effective temperature, and in most cases a faster relaxation of the distribution towards the thermalized one, as shown in the inset plot of the top panel of Fig.~\ref{fig:fermi_dist_and_trapping_analysis_A_lesser}, which plots the effective temperature extracted from the slope of $n_{\bf k}$ at $\omega=0$ (Eq.~\eqref{w_0_n_derivative_for_T}). Interestingly, different momenta at times $t \lesssim 10$ have significantly different distributions with different effective temperatures. On a timescale of $O(10)$, some distribution functions away from the Fermi level ($\mathbf{k}=\pm\frac{\pi}{2}$) approach the thermal ones. This suggests that the upturn/downturn in the $\pi$-ton spectra occurring near $t\sim 10$ (Fig.~\ref{fig:w_cut_2_down_1p5_1}) may be related to the $\mathbf{k}$-dependent relaxation of the distributions. The slow relaxation of the distribution at the Fermi level, which is expected from Fermi liquid theory, constitutes the main bottleneck in the thermalization of the $\pi$-ton.

In the quench down, where the heating effect is stronger, the distribution functions thermalize faster compared to the quench up (not shown). As a result, the prethermalization is less prominent. 

It is also possible to observe the prethermalization directly in the time dependence of the Green's functions. The bottom panel of 
Fig.~\ref{fig:fermi_dist_and_trapping_analysis_A_lesser} shows $\text{Im}\mathcal{G}^<(t,t+\Delta t')$ for $0\le \Delta t^{\prime} \le 20$. Again, the interaction is quenched from $U=1.5$ to $U=2$ at $T=0.05$. The inset shows the difference of the various time traces to the thermal result, that is $\text{Im}\mathcal{G}^<_\text{therm}(t,t+\Delta t')$ at $T_{\text{therm}}=0.0852$ and constant interaction $U=2$. 
These data confirm the slow relaxation to the time-dependence of the thermalized state, which is evident in the evolution of the distribution function. 

\section{Conclusion}
\label{sec:conclusion}

Using non-equilibrium calculations on the Kadanoff-Baym contour, we computed the longitudinal optical conductivity in the single-band half-filled Hubbard model after an interaction ramp or quench, considering single-ladder vertex corrections of the RPA $\pi$-ton type. First, we identified the relevant spectral features in  equilibrium systems with $U\lesssim \text{bandwidth}/2$ and, in agreement with previous studies related to the $\pi$-ton,\cite{kauch_pitons_2019,worm2020broadening,PhysRevB.103.104415_Simard_pi_ton,Bergeron_2011_optical_cond} found that a sharp spectral feature emerges at low energy in the optical conductivity when approaching the AFM phase boundary. We then studied the evolution of this intermediate-energy spectral feature, as well as the low-energy and high-energy features after quenches or ramps which increase or decrease $U$ in the vicinity of the AFM phase. While there was little qualitative difference between a quench and a (fast) ramp, the up and down quenches or ramps resulted in different dynamics. This is due to the effects of heating and the $U$-dependence of the AFM phase boundary. The up ramps or quenches result in states that are closer to the AFM boundary and hence feature strong $\pi$-ton type vertex corrections.  

Comparing time traces for different energy cuts, we found that the bubble contribution to the optical conductivity thermalizes fast after the ramp or quench. The corresponding time traces essentially feature a single strongly damped oscillation with a frequency that roughly matches the energy separation between the van Hove peaks in the DoS. In sharp contrast, the $\pi$-ton contribution to the conductivity exhibits a slower relaxation characterized by a long-lived nonthermal spectral distribution, especially at low energies. For the quench up, the main $\pi$-ton feature can transiently grow to values which significantly exceed the thermalized result. We have linked this observation to the prethermalization phenomenon that has been previously revealed in quenched, weakly interacting systems.\cite{Moeckel_2008} In particular, we have shown that while the spectral function thermalizes fast, the occupation after the quench can be distinctly nonthermal, especially for momenta near the Fermi level. Close to the Fermi energy, the relaxation of the occupation takes a long time, which due to the large velocity factors and the momentum difference $\mathbf{\pi}$ between the Fermi points, translates into a distinctly nonthermal $\pi$-ton contribution. 

Our analysis shows that ladder-type vertex corrections, which are prominent near the AFM phase boundary (or some other ordering instability with wave-vector $\pi$) have a significant effect on the optical properties in nonthermal, weakly-correlated Hubbard systems. In particular, prethermalization phenomena in these vertex corrections dominate the slow relaxation of the conductivity after a quench or other perturbation. 

\begin{acknowledgments}
We thank N. Bittner for helpful discussions. The calculations have been performed on the Beo05 cluster at the University of Fribourg. OS and PW acknolwedge support from ERC Consolidator Grant No. 724103.
\end{acknowledgments}


\appendix


\section{Single-ladder vertex corrections on the Kadanoff-Baym contour}
\label{sec:appendice:sl_vertex_corr_on_KB_contour}

\begin{figure}[ht]
  \centering
    \includegraphics[width=\linewidth]{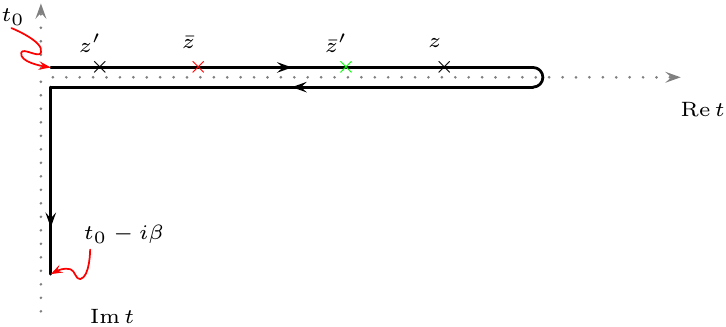}
      \caption{Kadanoff-Baym contour and the time arguments involved in the calculation of the single-ladder vertex corrections to the susceptibilities \eqref{eq:sus:non_eq_sl_KB_vertx_corr}. The integrated time variables are represented in colours, i.e red and green. In this particular situation, the configuration shows $\chi^{>}_{\text{sl}}(z,z^{\prime})$ with $\bar{z}\prec \bar{z}^{\prime}$, $\bar{z}\in \mathcal{C}_1$ and $\bar{z}^{\prime}\in \mathcal{C}_1$.}
  \label{fig:appendix:contour_example}
\end{figure}

In order to compute the retarded component of Eq.~\eqref{eq:sus:non_eq_sl_KB_vertx_corr}, one needs to work out both the lesser and greater components, implying that both contour parameters $z$ and $z^{\prime}$ in Eq.~\eqref{eq:sus:non_eq_sl_KB_vertx_corr} lie on the real-time axes. As a reminder, the lesser component $\chi_{\text{sl}}^{<}(z,z^{\prime})$ means that $z$ is encountered first following the contour ordering in Fig.~\ref{fig:appendix:contour_example} ($z^{\prime}\succ z$), whereas the greater component $\chi_{\text{sl}}^{>}(z,z^{\prime})$ means that $z^{\prime}$ is encountered first ($z^{\prime}\prec z$).

In what follows, we will write down the 9 different contributions to $\chi^{>}$ and $\chi^{<}$ arising from the internal integrals of the variables $\bar{z}$ and $\bar{z}^{\prime}$ over the different pieces of the contour. In the end, all these contributions to the greater/lesser components are summed up. The difference between the total greater and lesser components gives the retarded component. The Heaviside function on the contour $\theta^{\mathcal{C}}(z,z^{\prime})$ is defined such that $\theta^{\mathcal{C}}(z,z^{\prime}) = 1$ if $z\succ z^{\prime}$ and $\theta^{\mathcal{C}}(z,z^{\prime}) = 0$ if not. We also choose, without loss of generality, $z$ as the largest time on the real-time axis. The different contour functions that will show up are defined in Eq.~\eqref{eq:non_eq_components_cases}.

\begin{widetext}

\paragraph{$\bar{z}\in\mathcal{C}_1$ and $\bar{z}^{\prime}\in\mathcal{C}_1$}
\hfill \break
The contribution to the greater component ($z\succ z^{\prime}$) is

\begin{align}
\label{eq:contour_1:greater_part:a_gtr_b}
&\chi_{\text{sl}}^{>,\sigma,-\sigma}(\mathbf{q}; z,z^{\prime}) = -\int_{-\pi}^{\pi}\frac{\mathrm{d}^{D}\tilde{k}}{(2\pi)^{D}}\int_{-\pi}^{\pi}\frac{\mathrm{d}^{D}\bar{k}}{(2\pi)^{D}}\int^{z}_{t_0}\mathrm{d}\bar{z}\int_{t_0}^{z}\mathrm{d}\bar{z}^{\prime}\mathcal{G}^{>,\sigma}_{\tilde{\mathbf{k}}}(z,\bar{z})\theta^{\mathcal{C}}(z,\bar{z})\mathcal{G}^{<,\sigma}_{\tilde{\mathbf{k}}-\mathbf{q}}(\bar{z}^{\prime},z)\theta^{\mathcal{C}}(z,\bar{z}^{\prime})\times\notag\\
&\phantom{0} \left[\square^{>,\sigma,-\sigma}_{\tilde{\mathbf{k}}-\bar{\mathbf{k}}}(\bar{z},\bar{z}^{\prime})\theta^{\mathcal{C}}(\bar{z},\bar{z}^{\prime})+\square^{<,\sigma,-\sigma}_{\tilde{\mathbf{k}}-\bar{\mathbf{k}}}(\bar{z},\bar{z}^{\prime})\theta^{\mathcal{C}}(\bar{z}^{\prime},\bar{z})\right]\left[\mathcal{G}^{>,-\sigma}_{\bar{\mathbf{k}}}(\bar{z},z^{\prime})\theta^{\mathcal{C}}(\bar{z},z^{\prime})+\mathcal{G}^{<,-\sigma}_{\bar{\mathbf{k}}}(\bar{z},z^{\prime})\theta^{\mathcal{C}}(z^{\prime},\bar{z})\right]\times\notag\\
&\phantom{0} \left[\mathcal{G}^{<,-\sigma}_{\bar{\mathbf{k}}-\mathbf{q}}(z^{\prime},\bar{z}^{\prime})\theta^{\mathcal{C}}(\bar{z}^{\prime},z^{\prime})+\mathcal{G}^{>,-\sigma}_{\bar{\mathbf{k}}-\mathbf{q}}(z^{\prime},\bar{z}^{\prime})\theta^{\mathcal{C}}(z^{\prime},\bar{z}^{\prime})\right]\theta^{\mathcal{C}}(z,z^{\prime}),
\end{align}
while the contribution to the lesser component ($z^{\prime}\succ z$) reads

\begin{align}
\label{eq:contour_1:lesser_part:b_gtr_a}
&\chi_{\text{sl}}^{<,\sigma,-\sigma}(\mathbf{q}; z,z^{\prime}) = -\int_{-\pi}^{\pi}\frac{\mathrm{d}^{D}\tilde{k}}{(2\pi)^{D}}\int_{-\pi}^{\pi}\frac{\mathrm{d}^{D}\bar{k}}{(2\pi)^{D}}\int^{z}_{t_0}\mathrm{d}\bar{z}\int_{t_0}^{z}\mathrm{d}\bar{z}^{\prime}\mathcal{G}^{>,\sigma}_{\tilde{\mathbf{k}}}(z,\bar{z})\theta^{\mathcal{C}}(z,\bar{z})\mathcal{G}^{<,\sigma}_{\tilde{\mathbf{k}}-\mathbf{q}}(\bar{z}^{\prime},z)\theta^{\mathcal{C}}(z,\bar{z}^{\prime})\times\notag\\
&\phantom{0} \left[\square^{>,\sigma,-\sigma}_{\tilde{\mathbf{k}}-\bar{\mathbf{k}}}(\bar{z},\bar{z}^{\prime})\theta^{\mathcal{C}}(\bar{z},\bar{z}^{\prime})+\square^{<,\sigma,-\sigma}_{\tilde{\mathbf{k}}-\bar{\mathbf{k}}}(\bar{z},\bar{z}^{\prime})\theta^{\mathcal{C}}(\bar{z}^{\prime},\bar{z})\right]\mathcal{G}^{<,-\sigma}_{\bar{\mathbf{k}}}(\bar{z},z^{\prime})\theta^{\mathcal{C}}(z^{\prime},\bar{z})\mathcal{G}^{>,-\sigma}_{\bar{\mathbf{k}}-\mathbf{q}}(z^{\prime},\bar{z}^{\prime})\theta^{\mathcal{C}}(z^{\prime},\bar{z}^{\prime})\theta^{\mathcal{C}}(z^{\prime},z).
\end{align}
The Heaviside functions take care of the domain of integration. Subtracting Eq.~\eqref{eq:contour_1:lesser_part:b_gtr_a} from Eq.~\eqref{eq:contour_1:greater_part:a_gtr_b} yields the contribution to the retarded component. 
\newline

\paragraph{$\bar{z}\in \mathcal{C}_1$~and $\bar{z}^{\prime}\in \mathcal{C}_2$}
\hfill \break
The contribution to the greater component ($z\succ z^{\prime}$) is

\begin{align}
\label{eq:contour_2:greater_part:b_gtr_a}
&\chi_{\text{sl}}^{>,\sigma,-\sigma}(\mathbf{q}; z,z^{\prime}) = -\int_{-\pi}^{\pi}\frac{\mathrm{d}^{D}\tilde{k}}{(2\pi)^{D}}\int_{-\pi}^{\pi}\frac{\mathrm{d}^{D}\bar{k}}{(2\pi)^{D}}\int^{z}_{t_0}\mathrm{d}\bar{z}\int_{z}^{t_0}\mathrm{d}\bar{z}^{\prime}\mathcal{G}^{>,\sigma}_{\tilde{\mathbf{k}}}(z,\bar{z})\theta^{\mathcal{C}}(z,\bar{z})\mathcal{G}^{>,\sigma}_{\tilde{\mathbf{k}}-\mathbf{q}}(\bar{z}^{\prime},z)\theta^{\mathcal{C}}(\bar{z}^{\prime},z)\times\notag\\
&\phantom{0} \square^{<,\sigma,-\sigma}_{\tilde{\mathbf{k}}-\bar{\mathbf{k}}}(\bar{z},\bar{z}^{\prime})\theta^{\mathcal{C}}(\bar{z}^{\prime},\bar{z})\left[\mathcal{G}^{<,-\sigma}_{\bar{\mathbf{k}}}(\bar{z},z^{\prime})\theta^{\mathcal{C}}(z^{\prime},\bar{z})+\mathcal{G}^{>,-\sigma}_{\bar{\mathbf{k}}}(\bar{z},z^{\prime})\theta^{\mathcal{C}}(\bar{z},z^{\prime})\right]\mathcal{G}^{<,-\sigma}_{\bar{\mathbf{k}}-\mathbf{q}}(z^{\prime},\bar{z}^{\prime})\theta^{\mathcal{C}}(\bar{z}^{\prime},z^{\prime})\theta^{\mathcal{C}}(z,z^{\prime}),
\end{align}
while the contribution to the lesser component ($z^{\prime}\succ z$) reads

\begin{align}
\label{eq:contour_2:lesser_part:b_gtr_a}
&\chi_{\text{sl}}^{<,\sigma,-\sigma}(\mathbf{q}; z,z^{\prime}) = -\int_{-\pi}^{\pi}\frac{\mathrm{d}^{D}\tilde{k}}{(2\pi)^{D}}\int_{-\pi}^{\pi}\frac{\mathrm{d}^{D}\bar{k}}{(2\pi)^{D}}\int^{z}_{t_0}\mathrm{d}\bar{z}\int_{z}^{t_0}\mathrm{d}\bar{z}^{\prime}\mathcal{G}^{>,\sigma}_{\tilde{\mathbf{k}}}(z,\bar{z})\theta^{\mathcal{C}}(z,\bar{z})\mathcal{G}^{>,\sigma}_{\tilde{\mathbf{k}}-\mathbf{q}}(\bar{z}^{\prime},z)\theta^{\mathcal{C}}(\bar{z}^{\prime},z)\times\notag\\
&\phantom{0} \square^{<,\sigma,-\sigma}_{\tilde{\mathbf{k}}-\bar{\mathbf{k}}}(\bar{z},\bar{z}^{\prime})\theta^{\mathcal{C}}(\bar{z}^{\prime},\bar{z})\mathcal{G}^{<,-\sigma}_{\bar{\mathbf{k}}}(\bar{z},z^{\prime})\theta^{\mathcal{C}}(z^{\prime},\bar{z})\left[\mathcal{G}^{<,-\sigma}_{\bar{\mathbf{k}}-\mathbf{q}}(z^{\prime},\bar{z}^{\prime})\theta^{\mathcal{C}}(\bar{z}^{\prime},z^{\prime})+\mathcal{G}^{>,-\sigma}_{\bar{\mathbf{k}}-\mathbf{q}}(z^{\prime},\bar{z}^{\prime})\theta^{\mathcal{C}}(z^{\prime},\bar{z}^{\prime})\right]\theta^{\mathcal{C}}(z^{\prime},z).
\end{align}
The results for the case where $\bar{z}\in \mathcal{C}_2$~and $\bar{z}^{\prime}\in \mathcal{C}_1$ can be obtained from Eqs.~\eqref{eq:contour_2:greater_part:b_gtr_a} and \eqref{eq:contour_2:lesser_part:b_gtr_a} by swapping $\bar{z}$ and $\bar{z}^{\prime}$ from one branch to the other.
\newline

\paragraph{$\bar{z}\in \mathcal{C}_1$~and $\bar{z}^{\prime}\in \mathcal{C}_3$}
\hfill \break
The contribution to the greater component ($z\succ z^{\prime}$) is

\begin{align}
\label{eq:contour_3:greater_part:b_gtr_a}
&\chi_{\text{sl}}^{>,\sigma,-\sigma}(\mathbf{q}; z,z^{\prime}) = -\int_{-\pi}^{\pi}\frac{\mathrm{d}^{D}\tilde{k}}{(2\pi)^{D}}\int_{-\pi}^{\pi}\frac{\mathrm{d}^{D}\bar{k}}{(2\pi)^{D}}\int^{z}_{t_0}\mathrm{d}\bar{z}\int_{t_0}^{t_0-i\beta}\mathrm{d}\bar{z}^{\prime}\mathcal{G}^{>,\sigma}_{\tilde{\mathbf{k}}}(z,\bar{z})\theta^{\mathcal{C}}(z,\bar{z})\mathcal{G}^{\invneg,\sigma}_{\tilde{\mathbf{k}}-\mathbf{q}}(\bar{z}^{\prime},z)\times\notag\\
&\phantom{0} \square^{\neg,\sigma,-\sigma}_{\tilde{\mathbf{k}}-\bar{\mathbf{k}}}(\bar{z},\bar{z}^{\prime})\left[\mathcal{G}^{<,-\sigma}_{\bar{\mathbf{k}}}(\bar{z},z^{\prime})\theta^{\mathcal{C}}(z^{\prime},\bar{z})+\mathcal{G}^{>,-\sigma}_{\bar{\mathbf{k}}}(\bar{z},z^{\prime})\theta^{\mathcal{C}}(\bar{z},z^{\prime})\right]\mathcal{G}^{\neg,-\sigma}_{\bar{\mathbf{k}}-\mathbf{q}}(z^{\prime},\bar{z}^{\prime})\theta^{\mathcal{C}}(z,z^{\prime}),
\end{align}
while the contribution to the lesser component ($z^{\prime}\succ z$) reads

\begin{align}
\label{eq:contour_3:lesser_part:b_gtr_a}
&\chi_{\text{sl}}^{<,\sigma,-\sigma}(\mathbf{q}; z,z^{\prime}) = -\int_{-\pi}^{\pi}\frac{\mathrm{d}^{D}\tilde{k}}{(2\pi)^{D}}\int_{-\pi}^{\pi}\frac{\mathrm{d}^{D}\bar{k}}{(2\pi)^{D}}\int^{z}_{t_0}\mathrm{d}\bar{z}\int_{t_0}^{t_0-i\beta}\mathrm{d}\bar{z}^{\prime}\mathcal{G}^{>,\sigma}_{\tilde{\mathbf{k}}}(z,\bar{z})\theta^{\mathcal{C}}(z,\bar{z})\mathcal{G}^{\invneg,\sigma}_{\tilde{\mathbf{k}}-\mathbf{q}}(\bar{z}^{\prime},z)\times\notag\\
&\phantom{0} \square^{\neg,\sigma,-\sigma}_{\tilde{\mathbf{k}}-\bar{\mathbf{k}}}(\bar{z},\bar{z}^{\prime})\mathcal{G}^{<,-\sigma}_{\bar{\mathbf{k}}}(\bar{z},z^{\prime})\theta^{\mathcal{C}}(z^{\prime},\bar{z})\mathcal{G}^{\neg,-\sigma}_{\bar{\mathbf{k}}-\mathbf{q}}(z^{\prime},\bar{z}^{\prime})\theta^{\mathcal{C}}(z^{\prime},z).
\end{align} 
The results for $\bar{z}\in \mathcal{C}_3$~and $\bar{z}^{\prime}\in \mathcal{C}_1$ can be obtained by swapping $\bar{z}$ and $\bar{z}^{\prime}$ in Eqs.~\eqref{eq:contour_3:greater_part:b_gtr_a} and \eqref{eq:contour_3:lesser_part:b_gtr_a}.
\newline

\paragraph{$\bar{z}\in \mathcal{C}_2$~and $\bar{z}^{\prime}\in \mathcal{C}_2$}
\hfill \break

The contribution to the greater component ($z\succ z^{\prime}$) is

\begin{align}
\label{eq:contour_5:greater_part:a_gtr_b}
&\chi_{\text{sl}}^{>,\sigma,-\sigma}(\mathbf{q}; z,z^{\prime}) = -\int_{-\pi}^{\pi}\frac{\mathrm{d}^{D}\tilde{k}}{(2\pi)^{D}}\int_{-\pi}^{\pi}\frac{\mathrm{d}^{D}\bar{k}}{(2\pi)^{D}}\int^{t_0}_{z}\mathrm{d}\bar{z}\int_{z}^{t_0}\mathrm{d}\bar{z}^{\prime}\mathcal{G}^{<,\sigma}_{\tilde{\mathbf{k}}}(z,\bar{z})\theta^{\mathcal{C}}(\bar{z},z)\mathcal{G}^{>,\sigma}_{\tilde{\mathbf{k}}-\mathbf{q}}(\bar{z}^{\prime},z)\theta^{\mathcal{C}}(\bar{z}^{\prime},z)\times\notag\\
&\phantom{0} \left[\square^{>,\sigma,-\sigma}_{\tilde{\mathbf{k}}-\bar{\mathbf{k}}}(\bar{z},\bar{z}^{\prime})\theta^{\mathcal{C}}(\bar{z},\bar{z}^{\prime})+\square^{<,\sigma,-\sigma}_{\tilde{\mathbf{k}}-\bar{\mathbf{k}}}(\bar{z},\bar{z}^{\prime})\theta^{\mathcal{C}}(\bar{z}^{\prime},\bar{z})\right]\mathcal{G}^{>,-\sigma}_{\bar{\mathbf{k}}}(\bar{z},z^{\prime})\theta^{\mathcal{C}}(\bar{z},z^{\prime})\mathcal{G}^{<,-\sigma}_{\bar{\mathbf{k}}-\mathbf{q}}(z^{\prime},\bar{z}^{\prime})\theta^{\mathcal{C}}(\bar{z}^{\prime},z^{\prime})\theta^{\mathcal{C}}(z,z^{\prime}),
\end{align} 
while the contribution to the lesser component ($z^{\prime}\succ z$) reads

\begin{align}
\label{eq:contour_5:lesser_part:b_gtr_a}
&\chi_{\text{sl}}^{<,\sigma,-\sigma}(\mathbf{q}; z,z^{\prime}) = -\int_{-\pi}^{\pi}\frac{\mathrm{d}^{D}\tilde{k}}{(2\pi)^{D}}\int_{-\pi}^{\pi}\frac{\mathrm{d}^{D}\bar{k}}{(2\pi)^{D}}\int^{t_0}_{z}\mathrm{d}\bar{z}\int_{z}^{t_0}\mathrm{d}\bar{z}^{\prime}\mathcal{G}^{<,\sigma}_{\tilde{\mathbf{k}}}(z,\bar{z})\theta^{\mathcal{C}}(\bar{z},z)\mathcal{G}^{>,\sigma}_{\tilde{\mathbf{k}}-\mathbf{q}}(\bar{z}^{\prime},z)\theta^{\mathcal{C}}(\bar{z}^{\prime},z)\times\notag\\
&\phantom{0} \left[\square^{>,\sigma,-\sigma}_{\tilde{\mathbf{k}}-\bar{\mathbf{k}}}(\bar{z},\bar{z}^{\prime})\theta^{\mathcal{C}}(\bar{z},\bar{z}^{\prime})+\square^{<,\sigma,-\sigma}_{\tilde{\mathbf{k}}-\bar{\mathbf{k}}}(\bar{z},\bar{z}^{\prime})\theta^{\mathcal{C}}(\bar{z}^{\prime},\bar{z})\right]\left[\mathcal{G}^{>,-\sigma}_{\bar{\mathbf{k}}}(\bar{z},z^{\prime})\theta^{\mathcal{C}}(\bar{z},z^{\prime})+\mathcal{G}^{<,-\sigma}_{\bar{\mathbf{k}}}(\bar{z},z^{\prime})\theta^{\mathcal{C}}(z^{\prime},\bar{z})\right]\times\notag\\
&\phantom{0} \left[\mathcal{G}^{>,-\sigma}_{\bar{\mathbf{k}}-\mathbf{q}}(z^{\prime},\bar{z}^{\prime})\theta^{\mathcal{C}}(z^{\prime},\bar{z}^{\prime})+\mathcal{G}^{<,-\sigma}_{\bar{\mathbf{k}}-\mathbf{q}}(z^{\prime},\bar{z}^{\prime})\theta^{\mathcal{C}}(\bar{z}^{\prime},z^{\prime})\right]\theta^{\mathcal{C}}(z^{\prime},z).
\end{align}
\newline

\paragraph{$\bar{z}\in \mathcal{C}_2$~and $\bar{z}^{\prime}\in \mathcal{C}_3$}
\hfill \break

The contribution to the greater component ($z\succ z^{\prime}$) is

\begin{align}
\label{eq:contour_6:greater_part:b_gtr_a}
&\chi_{\text{sl}}^{>,\sigma,-\sigma}(\mathbf{q}; z,z^{\prime}) = -\int_{-\pi}^{\pi}\frac{\mathrm{d}^{D}\tilde{k}}{(2\pi)^{D}}\int_{-\pi}^{\pi}\frac{\mathrm{d}^{D}\bar{k}}{(2\pi)^{D}}\int^{t_0}_{z}\mathrm{d}\bar{z}\int_{t_0}^{t_0-i\beta}\mathrm{d}\bar{z}^{\prime}\mathcal{G}^{<,\sigma}_{\tilde{\mathbf{k}}}(z,\bar{z})\theta^{\mathcal{C}}(\bar{z},z)\mathcal{G}^{\invneg,\sigma}_{\tilde{\mathbf{k}}-\mathbf{q}}(\bar{z}^{\prime},z)\times\notag\\
&\phantom{0} \square^{\neg,\sigma,-\sigma}_{\tilde{\mathbf{k}}-\bar{\mathbf{k}}}(\bar{z},\bar{z}^{\prime})\mathcal{G}^{>,-\sigma}_{\bar{\mathbf{k}}}(\bar{z},z^{\prime})\theta^{\mathcal{C}}(\bar{z},z^{\prime})\mathcal{G}^{\neg,-\sigma}_{\bar{\mathbf{k}}-\mathbf{q}}(z^{\prime},\bar{z}^{\prime})\theta^{\mathcal{C}}(z,z^{\prime}),
\end{align} 

while the contribution to the lesser component ($z^{\prime}\succ z$) reads

\begin{align}
\label{eq:contour_6:lesser_part:b_gtr_a}
&\chi_{\text{sl}}^{<,\sigma,-\sigma}(\mathbf{q}; z,z^{\prime}) = -\int_{-\pi}^{\pi}\frac{\mathrm{d}^{D}\tilde{k}}{(2\pi)^{D}}\int_{-\pi}^{\pi}\frac{\mathrm{d}^{D}\bar{k}}{(2\pi)^{D}}\int^{t_0}_{z}\mathrm{d}\bar{z}\int_{t_0}^{t_0-i\beta}\mathrm{d}\bar{z}^{\prime}\mathcal{G}^{<,\sigma}_{\tilde{\mathbf{k}}}(z,\bar{z})\theta^{\mathcal{C}}(\bar{z},z)\mathcal{G}^{\invneg,\sigma}_{\tilde{\mathbf{k}}-\mathbf{q}}(\bar{z}^{\prime},z)\times\notag\\
&\phantom{0} \square^{\neg,\sigma,-\sigma}_{\tilde{\mathbf{k}}-\bar{\mathbf{k}}}(\bar{z},\bar{z}^{\prime})\left[\mathcal{G}^{<,-\sigma}_{\bar{\mathbf{k}}}(\bar{z},z^{\prime})\theta^{\mathcal{C}}(z^{\prime},\bar{z})+\mathcal{G}^{>,-\sigma}_{\bar{\mathbf{k}}}(\bar{z},z^{\prime})\theta^{\mathcal{C}}(\bar{z},z^{\prime})\right]\mathcal{G}^{\neg,-\sigma}_{\bar{\mathbf{k}}-\mathbf{q}}(z^{\prime},\bar{z}^{\prime})\theta^{\mathcal{C}}(z^{\prime},z).
\end{align}
The case where $\bar{z}\in \mathcal{C}_3$~and $\bar{z}^{\prime}\in \mathcal{C}_2$ can be deduced from Eqs.~\eqref{eq:contour_6:greater_part:b_gtr_a} and \eqref{eq:contour_6:lesser_part:b_gtr_a} by swapping the arguments.
\newline

\paragraph{$\bar{z}\in \mathcal{C}_3$~and $\bar{z}^{\prime}\in \mathcal{C}_3$}
\hfill \break
In this case, due to time translation invariance, the greater and lesser components are the same:

\begin{align}
\label{eq:contour_9:greater_part}
&\chi_{\text{sl}}^{>,\sigma,-\sigma}(\mathbf{q}; z,z^{\prime}) = -\int_{-\pi}^{\pi}\frac{\mathrm{d}^{D}\tilde{k}}{(2\pi)^{D}}\int_{-\pi}^{\pi}\frac{\mathrm{d}^{D}\bar{k}}{(2\pi)^{D}}\int^{t_0-i\beta}_{t_0}\mathrm{d}\bar{z}\int_{t_0}^{t_0-i\beta}\mathrm{d}\bar{z}^{\prime}\mathcal{G}^{\neg,\sigma}_{\tilde{\mathbf{k}}}(z,\bar{z})\mathcal{G}^{\invneg,\sigma}_{\tilde{\mathbf{k}}-\mathbf{q}}(\bar{z}^{\prime},z)\times\notag\\
&\phantom{0} \square^{M,\sigma,-\sigma}_{\tilde{\mathbf{k}}-\bar{\mathbf{k}}}(\bar{z}-\bar{z}^{\prime})\mathcal{G}^{\invneg,-\sigma}_{\bar{\mathbf{k}}}(\bar{z},z^{\prime})\mathcal{G}^{\neg,-\sigma}_{\bar{\mathbf{k}}-\mathbf{q}}(z^{\prime},\bar{z}^{\prime}) = \chi_{\text{sl}}^{<,\sigma,-\sigma}(\mathbf{q}; z,z^{\prime}).
\end{align}

\end{widetext}

\bibliography{Bibliography}
\end{document}